\documentclass[10pt,a4paper]{article}
\usepackage{eurosym}
\usepackage{graphicx}
\usepackage{amsmath}
\usepackage{amssymb}
\usepackage{latexsym}
\usepackage{color}
\usepackage{epstopdf}

\begin{document}

\thispagestyle{empty}

\begin{center}
\vspace{0.7cm}

{\Large \textbf{Generating Asymmetric Einstein-Podolsky-Rosen Steering
between Two movable Mirrors Exploiting Correlated-Emission Laser}}

\vspace{0.5cm}\textbf{Jamal El Qars$^{a,b}$, Ismail Essaoudi$^{a}$, and
Abdelmajid Ainane$^{a,c}$}

$^{a}$\textit{Laboratory of Materials Physics and Systems Modelling,
(LP2MS), Physics Department, Faculty of Sciences, Moulay Ismail University,
Meknes, Morocco}

$^{b}$\textit{Laboratory of Materials, Electrical Systems, Energy and
Environment, (LMS3E), Faculty of Applied Sciences, Ait-Melloul, Ibn Zohr
University, Agadir, Morocco}

$^{c}$\textit{Max-Planck-Institut f\"{u}r Physik Complexer Systeme, N\"{o}%
thnitzer Str. 38, Dresden 01187, Germany }

\vspace{0.9cm}

\vspace{0.9cm}\textbf{Abstract}
\end{center}

Quantum steering is a form of quantum correlation that exhibits an inherent
asymmetry, distinguishing it from entanglement and Bell nonlocality. It is
now understood that quantum steering plays a pivotal role in asymmetric
quantum information tasks. In this work, we propose a scheme to generate
asymmetric steering between two mechanical modes by transferring quantum
coherence from a correlated-emission laser. To accomplish this, we derive
quantum Langevin equations to describe the optomechanical coupling between
two cavity modes and two mechanical modes along with the master equation of
two-mode laser. By examining the case where the cavity modes scatter at the
anti-Stokes sidebands, we demonstrate that both two-way and one-way steering
can be achieved by adjusting the strength of the field driving the gain
medium of the laser. Furthermore, we show that the direction of one-way
steering can be controlled by varying the temperatures of the mechanical
baths or the strengths of the optomechanical couplings. Additionally, we
reveal that the directionality of one-way steering depends on the modes
fluctuation levels of the modes, with the mode exhibiting larger
fluctuations determining the direction. This highly controllable scheme
could potentially be realized with current technology, offering a promising
platform for implementing one-way quantum information tasks.

\section{Introduction}

The quantum nonlocality of entangled states was first highlighted by
Einstein, Podolsky, and Rosen (EPR) in 1935 \cite{epr}, where they argued
the incompatibility between local causality and the completeness of quantum
mechanics. By examining a general nonfactorizable pure state of two
particles, EPR demonstrated that a measurement performed on one particle
seemingly induces an apparent nonlocal perturbation on the wave-function of
other particle. For EPR, such nonlocal effects were unacceptable suggesting
the need for local hidden variables. The EPR paper prompted a significant
response later that year from Schr\"{o}dinger \cite{schro1,schro2} who
initially introduced the term \textit{steering} to analyse the EPR paradox,
and later extended the concept to mixed states. However, the formalization
of quantum steering was only fully developed in 2007 \cite{wiseman}.

In terms of the local hidden variable and local hidden state models, quantum
steering was rigorously defined in \cite{wiseman} as a type of quantum
correlation that allows an untrusted observer (e.g., Alice) to remotely
influence the state of another trusted observer (e.g., Bob) through
implementing local measurements \cite{uola}. Thus, steering is often
described as a one-sided device-independent scenario for verifying of
entanglement. In the hierarchy of nonseparable quantum correlations,
steering is positioned between entanglement \cite{horodecki} and Bell
nonlocality \cite{bell}, being strictly stronger than the former, but
strictly weaker than the later. In other words, not every entangled state
displays steering, and not every steerable state demonstrates Bell
nonlocality \cite{guhne}.

Unlike entanglement and Bell nonlocality, quantum steering is inherently
asymmetric. A bipartite state shared between Alice and Bob may be steerable
from Alice to Bob but not necessarily in the reverse direction, an asymmetry
demonstrated both theoretically \cite{bowles,qars1,qars2,yhlin1,yhlin2,Fancao} and experimentally
\cite{sun1,wolman,wolman2,Designolle,sun2}. This unique asymmetry has led to the recognition of
EPR steering as a key ingredient for one-sided device-independent quantum
key distribution (1SDI-QKD) \cite{branciard}, one-sided device-independent
quantum secret sharing \cite{armstrong}, universal one-way quantum computing
\cite{CLi}, secure quantum teleportation \cite{QHe}, and subchannel
discrimination \cite{watrous}. For these applications, effective
characterization of steerable states is essential, and considerable effort
has been dedicated to developing criteria for determining whether a
bipartite state is steerable \cite{reid,cavalcanti1,cavalcanti2} or for
quantifying the degree of steerability in such states \cite%
{watrous,Skrzypczyk,kogias}.

Observing quantum effects in macroscopic systems is of crucial importance,
as it can deepen our understanding of quantum nonlocality. Cavity
optomechanics, as an emerging field studying the strong coupling between
optical and mechanical degrees of freedom via radiation pressure \cite%
{aspelmayer}, offers a promising platform for exploring entanglement \cite%
{palomaki}, squeezing \cite{purdy}, optical bistability \cite{sete}, ground
state optical feedback cooling of fundamental mechanical modes \cite{sommer}%
, and even Schr\"{o}dinger's cat states \cite{cat}. Additionally,
nondegenerate three-level lasers---with a gain medium composed of
three-level atoms in a cascade configuration---have become an attractive
candidate for generating strong entangled photon pairs \cite{sz}. In these
lasers, quantum coherence between the upper and lower levels of a single
three-level atom serves as a fundamental component. This coherence can be
achieved by initially preparing the atoms in a coherent superposition of the
upper and lower levels (injected coherence process) \cite{vehn}, by coupling
the two levels using a strong external field (driven coherence process) \cite%
{xiong}, or by combining both methods \cite{ansari}.

Several theoretical proposals have shown that two spatially separated
movable mirrors can become entangled by transferring entanglement from a
correlated-emission laser in a doubly resonant cavity. For instance, Zhou
\textit{et al}. \cite{zhou} demonstrated that two optical modes can be
entangled through interaction with three-level atoms in a cascade
configuration, where atomic coherence is induced via the injected coherence
process. They essentially showed that this generated entanglement can then
be transferred to two movable mirrors via radiation pressure. In contrast,
using the driven coherence process, Ge \textit{et al}. \cite{nha} showed
that two movable mirrors can also be entangled in the strong optomechanical
coupling regime. Sete and Eleuch \cite{sete} employ both injected and driven
coherence process to achieve entanglement between two movable mirrors in the
adiabatic regime. Recently, various mechanisms to enhance the degree of
mechanical entanglement in a doubly resonant optomechanical cavity have been
analyzed, also within the adiabatic regime \cite{tesfa}.

Motivated partly by the recent achievements in cavity optomechanics \cite{groblacher,cohadon} and atom-cavity quantum electrodynamics \cite{QED1,QED2}; and partly by the considerable attention that has recently been paid to EPR steering as the essential resource for asymmetric quantum information precessing tasks \cite{uola}, we seek to extend the scope of discussion to investigate the possibility of implementing controllable
one-way steering between two mechanical modes by transferring quantum
coherence from a correlated-emission laser to these modes within a doubly
resonant optomechanical cavity. We emphasize that optomechanical steering has been explored in several scenarios without the mediation of
correlated-emission lasers \cite{liao,WDeng,HXG}. However, achieving practical one-way steering with flexible controllability remains a demanding issue. Varying some intrinsic physical parameters such as the optical decay rate or/and the mechanical damping rate is often employed as a main method to adjust the direction of one-way
steering \cite{QHe,WDeng,ZYang,SWu,HTan}. Unfortunately, such method, not only induces
additional losses and noises which affects the degree of the generated
steering, but also makes the experimental operations more complicated since optical and mechanical losses are mainly related to the surface roughness, impurities and defects of mirror materials \cite{aspelmayer}. Therefore, it is worth to further investigate the manipulation of the direction of one-way steering through external mechanism, rather than via intrinsic mechanism.

In this paper, we propose a scheme to generate asymmetric Gaussian quantum steering between two spatially separated mechanical modes, labelled as $m_{1}$ and $m_{2}$, by transferring quantum coherence from correlated emission laser to them. Importantly, we show that the direction of one-way steering between the modes $m_{1}$ and $m_{2}$ could be controlled via either the temperatures of the mechanical baths or the input powers of the cavity drive lasers, which is more practicable in experimental operations.

The remainder of this paper is organized as follows. In Section \ref{SecII},
we introduce our model which involves a nondegenerate three-level laser
interacting with two movable mirrors via radiation pressure force. In
Section \ref{SecIII}, using the master equation of two-mode laser coupled to
a vacuum reservoir, we derive the linearized quantum Langevin equations that
describe the optomechanical coupling between two cavity modes and two
mechanical modes. We then use these equations to obtain the covariance
matrix for the two mechanical modes in the steady-state. In Section \ref%
{SecIV}, using the measure proposed in \cite{kogias}, we quantify and
analyze the quantum steering between the two mechanical modes. Finally, in
Section \ref{SecV}, we present our conclusions.

\section{The Model and Hamiltonian of the Whole System \label{SecII}}

\begin{figure}[tbh]
\centerline{\includegraphics[width=7cm]{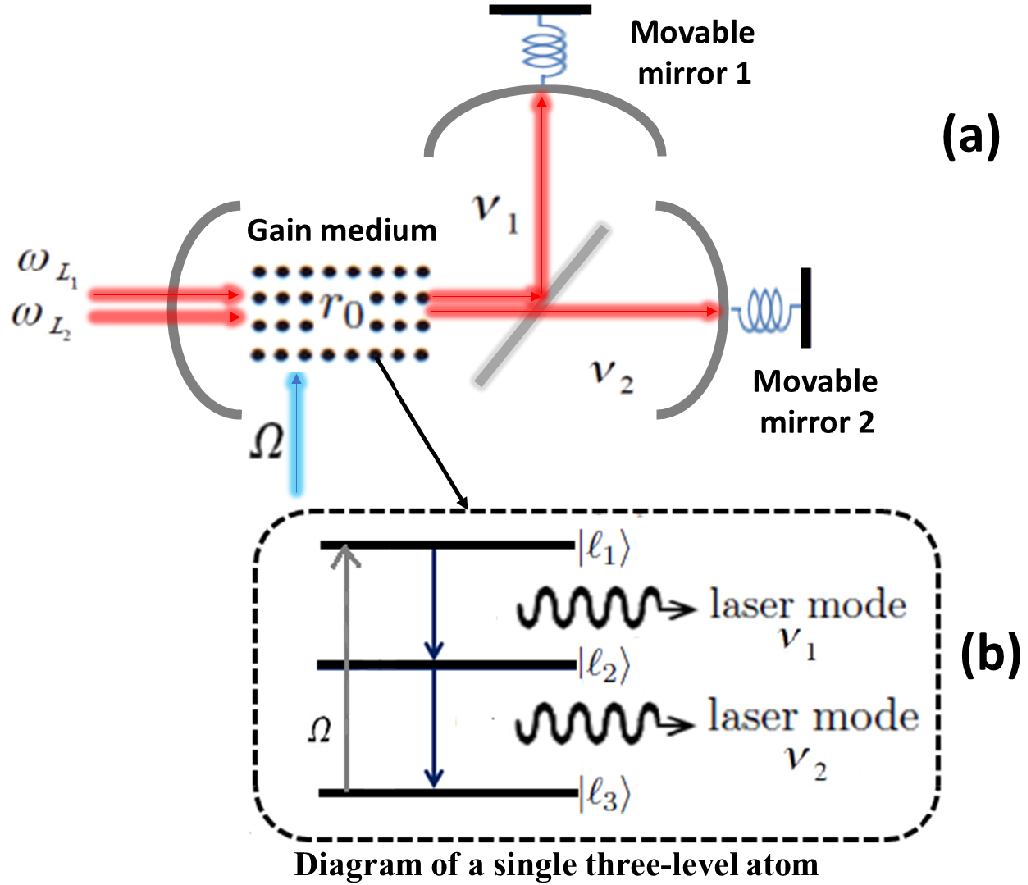}}
\caption{ (a) Scheme of generating quantum steering of two movable mirrors
inside a doubly resonant optomechanical cavity. The gain medium in the
cavity is driven by an external field of strength $\Omega $ and interacts
with two cavity modes of frequencies $\protect\nu _{1}$ and $\protect\nu %
_{2} $. The $j\text{th}$ cavity mode is driven by an external laser of
frequency $\protect\omega _{L_{j}}$, and coupled to its corresponding
movable mirror via radiation pressure force. (b) Schematic diagram of a
nondegenerate three-level laser coupled to a vacuum reservoir. $r_{0}$ is
the rate at which the atoms are injected into the cavity. $|\ell _{1}\rangle
$, $|\ell _{2}\rangle $, and $|\ell _{3}\rangle $ denote the upper,
intermediate, and lower energy levels of a single three-level atom,
respectively. The dipole-allowed transition $|\ell _{1}\rangle \rightarrow
|\ell _{2}\rangle $($|\ell _{2}\rangle \rightarrow |\ell _{3}\rangle $) is
assumed to be resonant with the cavity mode of frequency $\protect\nu _{1}$($%
\protect\nu _{2}$), while the dipole-forbidden transition $|\ell _{1}\rangle
\leftrightarrow |\ell _{3}\rangle $ can be induced by a resonant
semiclassical laser field of strength $\Omega $. $\protect\nu _{1}$($\protect%
\nu _{2}$) denote the frequency of the laser mode generated during the
transition $|\ell _{1}\rangle \rightarrow |\ell _{2}\rangle $($|\ell
_{2}\rangle \rightarrow |\ell _{3}\rangle $).}
\label{fig1}
\end{figure}

In Figure \ref{fig1}, we consider a nondegenerate three level laser with two
vibrating mirrors. The gain medium of the laser is an ensemble of
nondegenerate three-level atoms in a cascade configuration \cite{sz}, where
we use $|\ell _{1}\rangle $, $|\ell _{2}\rangle $, and $|\ell _{3}\rangle $
for denoting the upper, intermediate, and lower levels of a single atom,
respectively. The atoms are initially pumped into a doubly resonant cavity
at the lowest level $|\ell _{3}\rangle $ with a rate $r_{0}$, where the
dipole-allowed transition $|\ell _{1}\rangle \rightarrow |\ell _{2}\rangle $(%
$|\ell _{2}\rangle \rightarrow |\ell _{3}\rangle $) is assumed to be
resonant with a cavity mode $c_{1}(c_{2}$) having frequency $\nu _{1}$($\nu
_{2})$ and decay rate $\kappa _{1}(\kappa _{2}$). The $j$\textrm{th} cavity
mode is driven by an external coherent laser of frequency $\omega _{L_{j}}$,
power $\wp _{j}$, and amplitude $\epsilon _{j}=\sqrt{2\kappa _{j}\wp
_{j}/\hbar \omega _{L_{j}}}.$ The dipole-forbidden transition $|\ell
_{1}\rangle \leftrightarrow |\ell _{3}\rangle $ can be induced, for example,
by a strong magnetic field \cite{xiong}. The $j$\textrm{th} movable mirror
is modeled as a quantum-mechanical mode, labelled as $m_{j}$, having an
effective mass $\mu _{j}$, frequency $\omega _{m_{j}}$, damping rate $\gamma
_{m_{j}}$, and annihilation operator $b_{j}$.

In the interaction picture, under the rotating-wave approximation (RWA), the
Hamiltonian of the whole system writes $\mathcal{H}^{int}\mathcal{=H}%
_{af}^{int}+\mathcal{H}_{ac}^{int}+\mathcal{H}_{op}^{int}$, where $\mathcal{H%
}_{af}^{int}=\mathrm{i}\hbar \sum\limits_{j=1}^{2}\varsigma _{j}\left(
c_{j}|\ell _{j}\rangle \langle \ell _{j+1}|-|\ell _{j+1}\rangle \langle \ell
_{j}|c_{j}^{\dag }\right) $ describes the interaction between a single
three-level atom and two modes of the cavity field, with $\varsigma _{1}$($%
\varsigma _{2}$) being the coupling strength between the transition $|\ell
_{1}\rangle \rightarrow |\ell _{2}\rangle $($|\ell _{2}\rangle \rightarrow
|\ell _{3}\rangle $) and the cavity mode of annihilation operator $%
c_{1}(c_{2})$ \cite{xiong}. Without affecting the generality of our study,
we take identical spontaneous decay rates for both transitions $|\ell
_{1}\rangle \rightarrow |\ell _{2}\rangle $ and $|\ell _{2}\rangle
\rightarrow |\ell _{3}\rangle $, i.e., $\gamma _{1,2}=\gamma $, and
identical coupling transitions, i.e., $\varsigma _{1,2}=\varsigma $. The
Hamiltonian $\mathcal{H}_{ac}^{int}=\mathrm{i}\hbar \frac{\Omega }{2}(|\ell
_{1}\rangle \langle \ell _{3}|-|\ell _{3}\rangle \langle \ell _{1}|)$
corresponds to the generation of atomic coherence between the lower level $%
|\ell _{1}\rangle $ and the upper level $|\ell _{3}\rangle $ via an external
driving field of strength $\Omega $. The optomechanical Hamiltonian $%
\mathcal{H}_{op}^{int}$ is given by
\begin{equation}
\mathcal{H}_{op}^{int}=\sum\limits_{j=1}^{2}\left[ \hbar \Delta
_{j}c_{j}^{\dag }c_{j}+\hbar \omega _{m_{j}}b_{j}^{\dag }b_{j}-\hbar
g_{j}c_{j}^{\dag }c_{j}(b_{j}^{\dag }+b_{j})+\mathrm{i}\hbar \epsilon
_{j}(c_{j}^{\dag }e^{i\Delta _{j}^{\prime }}-c_{j}e^{-i\Delta _{j}^{\prime
}})\right] ,  \label{op}
\end{equation}%
where $\Delta _{j}$ is the $j\mathrm{th}$ laser detuning induced by
radiation pressure force \cite{aspelmayer}. The optomechanical coupling
strength between the $j\mathrm{th}$ cavity mode and its corresponding
mechanical mode is given by $g_{j}=\left( \nu _{j}/l_{j}\right) \sqrt{\hbar
/\mu _{_{j}}\omega _{m_{j}}}$, with $l_{j}$ being the $j\text{th}$ cavity
length at equilibrium. Finally, $\Delta _{j}^{\prime }=\nu _{j}-\Delta
_{j}-\omega _{L_{j}}$ denotes the effective detuning of the $j\mathrm{th}$
cavity-driving field.

The dynamics of the reduced density operator $\rho _{o_{1}o_{2}}\equiv \rho $
of two-mode laser $o_{1}$ and $o_{2}$ generated during the first and second
transitions respectively, can be described by the master equation \cite%
{qars2,Louissel}
\begin{equation}
\frac{d\rho }{dt}=\frac{-\text{i}}{\hbar }\mathrm{Tr}_{\mathrm{sa}}[\mathcal{%
H}_{af}^{int}+\mathcal{H}_{ac}^{int},\rho _{(\mathrm{sa},o_{1}o_{2})}]+\sum%
\limits_{j=1,2}\kappa _{j}\mathcal{L}\left[ c_{j}\right] \rho ,  \label{E4}
\end{equation}%
where $\rho _{(\mathrm{sa},o_{1}o_{2})}$ is the density operator describing
the two-mode laser together with a single three-level atom, and $\text{Tr}_{%
\text{sa}}$ denotes the partial trace over the state of a single atom. The
Lindblad operator $\mathcal{L}\left[ c_{j}\right] \rho =2c_{j}\rho
c_{j}^{\dag }-c_{j}^{\dag }c_{j}\rho -\rho c_{j}^{\dag }c_{j}$ is added to
take into account the coupling between the laser modes and the vacuum
reservoir \cite{milburn}. In the good cavity limit where $\kappa _{j}\ll
\gamma _{j}$, the atoms reach their steady state much faster than the cavity
modes, then the dynamics of the atoms can be eliminated adiabatically. Under
such condition, El Qars \cite{qars2} has derived the master equation of the
reduced density operator for two-mode laser coupled to two-mode squeezed
vacuum reservoir. Following the standard methods of laser theory \cite%
{Louissel,Sargent}, one can show that Equation (\ref{E4}) would be (details
of the calculations are given in Appendix B)
\begin{eqnarray}
\frac{d\rho }{dt} &=&\Xi _{11}\left( 2c_{1}^{\dag }\rho
c_{1}-c_{1}c_{1}^{\dag }\rho -\rho c_{1}c_{1}^{\dag }\right) +\Xi
_{22}\left( 2c_{2}\rho c_{2}^{\dag }-c_{2}^{\dag }c_{2}\rho -\rho
c_{2}^{\dag }c_{2}\right)  \notag \\
&&-\Xi _{12}c_{1}c_{2}\rho -\Xi _{21}\rho c_{1}c_{2}+\left( \Xi _{12}+\Xi
_{21}\right) c_{2}\rho c_{1}  \notag \\
&&-\Xi _{12}\rho c_{1}^{\dag }c_{2}^{\dag }-\Xi _{21}c_{1}^{\dag
}c_{2}^{\dag }\rho +\left( \Xi _{12}+\Xi _{21}\right) c_{1}^{\dag }\rho
c_{2}^{\dag }  \notag \\
&&+\kappa _{1}\left( 2c_{1}\rho c_{1}^{\dag }-c_{1}^{\dag }c_{1}\rho -\rho
c_{1}^{\dag }c_{1}\right) +\kappa _{2}\left( 2c_{2}\rho c_{2}^{\dag
}-c_{2}^{\dag }c_{2}\rho -\rho c_{2}^{\dag }c_{2}\right) ,  \label{hn}
\end{eqnarray}%
where
\begin{eqnarray}
\Xi _{11} &=&\frac{3\mathcal{A}}{8}\frac{\Omega ^{2}\gamma ^{2}}{(\Omega
^{2}+\gamma ^{2})(\frac{\Omega ^{2}}{4}+\gamma ^{2})},\text{ }\Xi _{22}=%
\frac{\mathcal{A}}{2}\frac{\gamma ^{2}}{(\Omega ^{2}+\gamma ^{2})},
\label{co1} \\
\Xi _{12} &=&-\frac{\mathcal{A}}{2}\frac{\Omega \gamma }{(\Omega ^{2}+\gamma
^{2})},\text{\ \ and \ \ \ }\Xi _{21}=\frac{\mathcal{A}}{8}\frac{\Omega
\gamma (\Omega ^{2}-2\gamma ^{2})}{(\Omega ^{2}+\gamma ^{2})(\frac{\Omega
^{2}}{4}+\gamma ^{2})},  \label{co2}
\end{eqnarray}%
with $\mathcal{A}=2r_{0}\varsigma ^{2}/\gamma ^{2}$ being the linear gain
coefficient that quantifies the rate at which the atoms are injected into
the cavity \cite{sz}. In Equation (\ref{hn}) the term proportional to $\Xi
_{11}$($\Xi _{22}$) corresponds to the gain(loss) of the first(second) laser
mode, while the next six terms traduce the coupling between the two emitted
laser modes due to atomic coherence induced by the driven field of strength $%
\Omega $. We emphasize that such terms are responsible for the creation of
quantum correlations between the two generated laser modes \cite{sz}. In
this work, we will exploit these correlations to produce asymmetric Gaussian
quantum steering between two non-interacting movable mirrors. Finally, the
two last terms in Equation (\ref{hn}) proportional to $\kappa _{1}$ and $%
\kappa _{2}$ traduce the damping of the two-mode laser in the vacuum
reservoir.

\section{Quantum Langevin Equations for the Optomechanical Subsystem}

\label{SecIII}

Now we derive the quantum Langevin equations for the two cavity modes $%
c_{1,2}$ and the two mechanical modes $m_{1,2}$. For this, we assume the
coupling between a single atom and the cavity modes to be much stronger than
the optomechanical coupling which allows us to obtain the quantum Langevin
equations for the atom--cavity subsystem and the optomechanical subsystem
separately. The quantum Langevin equations for the cavity modes could be
derived from Equation (\ref{hn}) using the formula $\frac{d}{dt}\langle
\mathcal{O}\rangle =\mathrm{Tr}\left( \frac{d\rho }{dt}\mathcal{O}\right) $
for $\mathcal{O}_{j}\equiv c_{j}$ and removing the bracket $\langle \mathcal{%
.}\rangle $ from the resulting equations, and next adding appropriate noise
operators $\digamma _{j}$ with zero mean value ($\langle \digamma
_{j}\rangle =0$) . Therefore, we get (details of the calculations are given
in Appendix C)%
\begin{eqnarray}
\partial _{t}c_{1} &=&-\left( \kappa _{1}-\Xi _{11}\right) c_{1}+\Xi
_{12}c_{2}^{\dag }+\digamma _{1},  \label{E2} \\
\partial _{t}c_{2} &=&-\left( \kappa _{2}+\Xi _{22}\right) c_{2}-\Xi
_{21}c_{1}^{\dag }+\digamma _{2},  \label{E3}
\end{eqnarray}%
where the time-domain correlation functions of the noise operators $\digamma
_{1}$ and $\digamma _{2}$ can be obtained from Einstein's relation \cite%
{einstein}
\begin{equation}
2\langle \mathcal{D}_{\mathcal{O}_{1}\mathcal{O}_{2}}\rangle =\frac{d\langle
\mathcal{O}_{1}\mathcal{O}_{2}\rangle }{dt}-\langle \left( \mathcal{\dot{O}}%
_{1}-\digamma _{\mathcal{O}_{1}}\right) \mathcal{O}_{2}\rangle -\langle
\mathcal{O}_{1}\left( \mathcal{\dot{O}}_{2}-\digamma _{\mathcal{O}%
_{2}}\right) \rangle ,  \label{ein}
\end{equation}%
with $\langle \mathcal{D}_{\mathcal{O}_{1}\mathcal{O}_{2}}\rangle $ being
the diffusion coefficient for $\mathcal{O}_{j}\equiv c_{j}$, and $\digamma _{%
\mathcal{O}_{1}}$ and $\digamma _{\mathcal{O}_{2}}$ belong to their
corresponding noise operators. Using Equation (\ref{ein}) and the equations
for the second-order moments of the laser modes operators $c_{j}$ together
with $\langle \digamma _{\mathcal{O}_{1}}(t)\digamma _{\mathcal{O}%
_{2}}(t^{\prime })\rangle =2\langle \mathcal{D}_{\mathcal{O}_{1}\mathcal{O}%
_{2}}\rangle \delta (t-t^{\prime })$, the non-vanishing correlation
functions for the operators $\digamma _{1}$ and $\digamma _{2}$ are

\begin{eqnarray}
\langle \digamma _{1}^{\dag }(t)\digamma _{1}(t^{\prime })\rangle &=&2\Xi
_{11}\delta (t-t^{\prime }),  \label{ein1} \\
\langle \digamma _{1}(t)\digamma _{1}^{\dag }(t^{\prime })\rangle &=&2\kappa
_{1}\delta (t-t^{\prime }),  \label{ein2} \\
\langle \digamma _{2}(t)\digamma _{2}^{\dag }(t^{\prime })\rangle &=&2\left(
\kappa _{2}+\Xi _{22}\right) \delta (t-t^{\prime }),  \label{ein3} \\
\langle \digamma _{1}^{\dag }(t)\digamma _{2}^{\dag }(t^{\prime })\rangle
&=&\langle \digamma _{2}(t)\digamma _{1}(t^{\prime })\rangle =-2\left( \Xi
_{12}+\Xi _{21}\right) \delta (t-t^{\prime }).  \label{ein4}
\end{eqnarray}

We emphasize that since the quantum system at hand is intrinsically open, consequently it is inevitably affected by the environmental noises. Then, the quantum noise operators $\digamma_{1}$ and $\digamma_{2}$, appeared in Equations (\ref{E2}) and (\ref{E3}), are added to take into account the effect of the noise resulting from the coupling between the laser modes and the vacuum reservoir.

Now, a more general dynamics of the laser modes can be obtained by adding the contribution of the optomechanical Hamiltonian (\ref{op}) in Equations (\ref{E2}) and (\ref{E3}). To do this, we add $\frac{1}{i\hbar}[c_{1},\mathcal{H}^{int}_{op}]$ and $\frac{1}{i\hbar}[c_{2},\mathcal{H}^{int}_{op}]$, respectively, in the right hand side of Equations (\ref{E2}) and (\ref{E3}). Then, we get
\begin{eqnarray}
\partial _{t}c_{1} &=&-\left( \kappa _{1}-\Xi _{11}+i\Delta _{1}\right)
c_{1}+\Xi _{12}c_{2}^{\dag }+ig_{1}c_{1}(b_{1}^{\dag }+b_{1})+\epsilon
_{1}e^{i\Delta _{1}^{\prime }}+\digamma _{1},  \label{c1} \\
\partial _{t}c_{2} &=&-\left( \kappa _{2}+\Xi _{22}+i\Delta _{2}\right)
c_{2}-\Xi _{21}c_{1}^{\dag }+ig_{2}c_{2}(b_{2}^{\dag }+b_{2})+\epsilon
_{2}e^{i\Delta _{2}^{\prime }}+\digamma _{2}.  \label{c2}
\end{eqnarray}

The two mechanical modes are also affected by damping and noise processes, due to the fact that each of them interacts with its own thermal bath. Therefore, their dynamics can be obtained by adopting the quantum Langevin equations treatment, in which the Heisenberg equation associated with the Hamiltonian (\ref{op}) is supplemented with mechanical damping and Brownian noise terms, i.e., $\partial_{t}b_{j}=\frac{1}{i\hbar}[b_{j},\mathcal{H}^{int}_{op}]-\gamma_{m_{j}}b_{j}+\sqrt{2\gamma_{m_{j}}}\zeta_{j}$ \cite{Barza}. Then, we obtain

\begin{eqnarray}
\partial _{t}b_{j} &=&-\left( \gamma _{m_{j}}+i\omega _{m_{j}}\right)
b_{j}+ig_{j}c_{j}^{\dag }c_{j}+\sqrt{2\gamma _{m_{j}}}\zeta _{j},\text{ for }%
j=1,2\text{,}  \label{mj}
\end{eqnarray}
where $\zeta _{j}$ denotes the Brownian noise operator, with zero-mean
value, that affects the $j\mathrm{th}$ mechanical mode at temperature $T_{j}$%
. In general, the operator $\zeta _{j}$ is not $\delta $-correlated,
exhibiting a non-Markovian correlation property between two instants $t$ and
$t^{\prime }$ \cite{genes}. However, large mechanical quality factor $%
\mathcal{Q}_{m_{j}}=\omega _{m_{j}}/\gamma _{m_{j}}\gg 1$ allows us
recovering the Markovian process. Then, we have \cite{benguirria}
\begin{eqnarray}
\langle \zeta _{j}^{\dag }(t)\zeta _{j}(t^{\prime })\rangle &=&n_{\text{th}%
,j}\delta (t-t^{\prime })\text{ for \ }j=1,2,  \label{ben1} \\
\langle \zeta _{j}(t)\zeta _{j}^{\dag }(t^{\prime })\rangle &=&\left( n_{%
\text{th},j}+1\right) \delta (t-t^{\prime }),  \label{ben2}
\end{eqnarray}%
where $n_{\text{th},j}=(e^{\hbar \omega _{m_{j}}/k_{B}T_{j}}-1)^{-1}$ is the
$j\mathrm{th}$ mean thermal phonon number and $k_{B}$ is the Boltzmann
constant. We emphasize that non-Markovian evolution of EPR steering has been
recently observed experimentally in \cite{guo}.

Due to the quadratic terms $c_{j}^{\dag }c_{j}$, $c_{j}b_{j}^{\dag }$ and $%
c_{j}b_{j}$, the nonlinear quantum Langevin equations [(\ref{c1})-(\ref{mj}%
)] cannot be solved analytically. However, assuming bright pump fields, one
can linearize the dynamics around the steady state by adopting standard
method of quantum optics \cite{genes}. To realise this, we first choose a
rotating frame defined as $\tilde{c}_{j}=c_{j}e^{-i\Delta _{j}^{\prime }t}$,
next we write each Heisenberg operator $\mathcal{O}_{j}\equiv \tilde{c}%
_{j},b_{j}$ as a complex steady state value $\langle \mathcal{O}_{j}\rangle $
plus a small fluctuation $\delta \mathcal{O}_{j}$ with vanishing-mean value,
i.e., $\mathcal{O}_{j}=\langle \mathcal{O}_{j}\rangle +\delta \mathcal{O}%
_{j} $. In the choosing rotating frame, the obtained equations for the
fluctuations $\delta \tilde{c}_{1}$ and $\delta \tilde{c}_{2}$ as well as
the mean values $\langle \tilde{c}_{1}\rangle $ and $\langle \tilde{c}%
_{2}\rangle $ will be coupled through terms proportional to $\Xi
_{12}e^{-i(\Delta _{1}^{\prime }+\Delta _{2}^{\prime })t}$ and $\Xi
_{21}e^{-i(\Delta _{1}^{\prime }+\Delta _{2}^{\prime })t}$. Then, by
performing the RWA that allows us to drop the highly oscillating terms with
the factor $e^{-i(\Delta _{1}^{\prime }+\Delta _{2}^{\prime })t}$, solely in
the quantum Langevin equations of $\langle \tilde{c}_{j}\rangle $, and next
setting the time derivatives in the resulting equations, we obtain $\langle
b_{j}\rangle ^{\ast }+\langle b_{j}\rangle =\frac{\left\vert \langle \tilde{c%
}_{j}\rangle \right\vert ^{2}}{\omega _{m_{j}}}$ and $\langle \tilde{c}%
_{j}\rangle =\frac{\epsilon _{j}}{\kappa _{j}+i\Delta _{j}^{\prime
}+(-1)^{j}\Xi _{jj}}$, where $\Delta _{j}^{\prime }=\nu _{j}-\omega
_{L_{j}}-g_{j}\left( \langle b_{j}\rangle ^{\ast }+\langle b_{j}\rangle
\right) $ denotes the effective detuning of the $j$\textrm{th} cavity mode
frequency. Hence, it follows that the mean frequency shift $\Delta _{j}$,
introduced in Equation (\ref{op}), is given by $\Delta _{j}=g_{j}\left(
\langle b_{j}\rangle ^{\ast }+\langle b_{j}\rangle \right) .$

Now, transforming back to the original rotating frame, i.e., $\delta
c_{j}=\delta \tilde{c}_{j}e^{i\Delta _{j}^{\prime }t}$ and introducing $%
\delta \tilde{b}_{j}=\delta b_{j}e^{-i\omega _{j}t}$, one has
\begin{eqnarray}
\partial _{t}\delta c_{1} &=&-\bar{\kappa}_{1}\delta c_{1}+\Xi _{12}\delta
c_{2}^{\dag }+iG_{1}\left( \delta \tilde{b}_{1}e^{-i\left( \omega
_{m_{1}}-\Delta _{1}^{\prime }\right) t}+\delta \tilde{b}_{1}^{\dag
}e^{i\left( \omega _{m_{1}}+\Delta _{1}^{\prime }\right) t}\right) +\digamma
_{1},  \label{c11} \\
\partial _{t}\delta c_{2} &=&-\bar{\kappa}_{2}\delta c_{2}-\Xi _{21}\delta
c_{1}^{\dag }+iG_{2}\left( \delta \tilde{b}_{2}e^{-i\left( \omega
_{m_{2}}-\Delta _{2}^{\prime }\right) t}+\delta \tilde{b}_{2}^{\dag
}e^{i\left( \omega _{m_{2}}+\Delta _{2}^{\prime }\right) t}\right) +\digamma
_{2},  \label{c22} \\
\partial _{t}\delta \tilde{b}_{j} &=&-\gamma _{m_{j}}\delta \tilde{b}%
_{j}+iG_{j}^{\ast }\delta c_{j}e^{i\left( \omega _{m_{j}}-\Delta
_{j}^{\prime }\right) t}+iG_{j}\delta c_{j}^{\dag }e^{i\left( \omega
_{m_{j}}+\Delta _{j}^{\prime }\right) t}+\sqrt{2\gamma _{m_{j}}}\tilde{\zeta}%
_{j},\text{ for }j=1,2,  \label{mjj}
\end{eqnarray}%
where $\bar{\kappa}_{j}=\kappa _{j}+(-1)^{j}\Xi _{jj}$, $\tilde{\zeta}%
_{j}=\zeta _{j}e^{i\omega _{m_{j}}t}$, and $G_{j}=g_{j}\langle \tilde{c}%
_{j}\rangle $ being the effective many-photon optomechanical coupling \cite%
{aspelmayer}. Notice that one can easily verify that the operator $\zeta
_{j} $ fulfills the same correlation properties given by Equations (\ref%
{ben1}) and (\ref{ben2}).

When the input lasers are scattered at the anti-Stokes sidebands, i.e., $%
\Delta _{j}^{\prime }=+\omega _{m_{j}}$, the operators $\delta c_{j}$ and $%
\delta \tilde{b}_{j}$ will be coupled via a beam-splitter-like process \cite%
{genes}, which is shown to be a very stable regime \cite{nha}. Importantly,
although there is no direct interaction between the two mechanical modes $%
m_{1}$ and $m_{2}$, entanglement between them could be generated via
transferring quantum correlations from the correlated-emission laser to the
state of the two movable mirrors. Indeed, under the condition $\kappa
_{j}\gg \gamma _{m_{j}}$, which is considered in this work, the effective
coupling between the modes $m_{1}$ and $m_{2}$ can be obtained by
eliminating adiabatically the dynamics of the cavity modes from Equations (%
\ref{c11}) and (\ref{c22}) and replacing into Equation (\ref{mjj}). Then, in
the resulting equations, the two mechanical modes will be coupled to each
other via a parametric down conversion process which leads to entanglement
between them \cite{genes}.

Using $\Delta _{j}^{\prime }=+\omega _{m_{j}}$ for $j=1,2$ in Equations [(%
\ref{c11})-(\ref{mjj})], and assuming that $\omega _{m_{j}}\gg \kappa _{j}$,
therefore the RWA allows us to neglect the terms rapidly oscillating at $%
+2\omega _{m_{j}}$ \cite{qars1}, hence we get
\begin{eqnarray}
\partial _{t}\delta c_{1} &=&-\bar{\kappa}_{1}\delta c_{1}+\Xi _{12}\delta
c_{2}^{\dag }+iG_{1}\delta \tilde{b}_{1}+\digamma _{1},  \label{c33} \\
\partial _{t}\delta c_{2} &=&-\bar{\kappa}_{2}\delta c_{2}-\Xi _{21}\delta
c_{1}^{\dag }+iG_{2}\delta \tilde{b}_{2}+\digamma _{2},  \label{c34} \\
\partial _{t}\delta \tilde{b}_{j} &=&-\gamma _{m_{j}}\delta \tilde{b}%
_{j}+iG_{j}^{\ast }\delta c_{j}+\sqrt{2\gamma _{m_{j}}}\tilde{\zeta}_{j}%
\text{ for }j=1,2  \label{m33}
\end{eqnarray}

Using the dimenssionless position and momentum fluctuations of the $j\mathrm{%
th}$ mechanical(cavity) mode $\delta \tilde{q}_{m_{j}}=(\delta \tilde{b}%
_{j}^{\dag }+\delta \tilde{b}_{j})/\sqrt{2}$ and $\delta \tilde{p}%
_{m_{j}}=i(\delta \tilde{b}_{j}^{\dag }-\delta \tilde{b}_{j})/\sqrt{2}$ $%
\Big(\delta q_{c_{j}}=(\delta c_{j}^{\dag }+\delta c_{j})/\sqrt{2}$ and $%
\delta p_{c_{j}}=i(\delta c_{j}^{\dag }-\delta c_{j})/\sqrt{2}\Big)$ with
the input noises $\tilde{q}_{m_{j}}^{in}=(\tilde{\zeta}_{j}^{\dagger }+%
\tilde{\zeta}_{j})/\sqrt{2}$ and $\tilde{p}_{m_{j}}^{in}=i(\tilde{\zeta}%
_{j}^{\dagger }-\tilde{\zeta}_{j})/\sqrt{2}$ $\Big(q_{c_{j}}^{in}=(\digamma
_{j}^{\dag }+\digamma _{j})/\sqrt{2}$ and $p_{c_{j}}^{in}=i(\digamma
_{j}^{\dag }-\digamma _{j})/\sqrt{2}\Big)$, and Equations [(\ref{c33})-(\ref%
{m33})], we obtain the linear quantum Langevin equations in the RWA in a
compact form as $\partial _{t}\mathcal{U}=\mathcal{KU}+\mathcal{N}$, with $%
\mathcal{U}^{\mathrm{T}}=(\delta \tilde{q}_{m_{1}},\delta \tilde{p}%
_{m_{1}},\delta \tilde{q}_{m_{2}},\delta \tilde{p}_{m_{2}},\delta
q_{c_{1}},\delta q_{c_{1}},\delta q_{c_{2}},\delta q_{c_{2}})$, $\mathcal{N}%
^{\mathrm{T}}=(\sqrt{2\gamma _{m_{1}}}\tilde{q}_{m_{1}}^{in},\sqrt{2\gamma
_{m_{1}}}\tilde{p}_{m_{1}}^{in},\sqrt{2\gamma _{m_{2}}}\tilde{q}%
_{m_{2}}^{in},\sqrt{2\gamma _{m_{2}}}\tilde{p}%
_{m_{2}}^{in},q_{c_{1}}^{in},p_{c_{1}}^{in},q_{c_{2}}^{in},p_{c_{2}}^{in})$
and the kernel $\mathcal{K}$ is given by%
\begin{equation}
\mathcal{K=}\left(
\begin{array}{cccccccc}
-\gamma _{m_{1}} & 0 & 0 & 0 & 0 & -G_{1} & 0 & 0 \\
0 & -\gamma _{m_{1}} & 0 & 0 & G_{1} & 0 & 0 & 0 \\
0 & 0 & -\gamma _{m_{2}} & 0 & 0 & 0 & 0 & -G_{2} \\
0 & 0 & 0 & -\gamma _{m_{2}} & 0 & 0 & G_{2} & 0 \\
0 & -G_{1} & 0 & 0 & -\bar{\kappa}_{1} & 0 & \Xi _{12} & 0 \\
G_{1} & 0 & 0 & 0 & 0 & -\bar{\kappa}_{1} & 0 & -\Xi _{12} \\
0 & 0 & 0 & -G_{2} & -\Xi _{21} & 0 & -\bar{\kappa}_{2} & 0 \\
0 & 0 & G_{2} & 0 & 0 & \Xi _{21} & 0 & -\bar{\kappa}_{2}%
\end{array}%
\right) ,  \label{e13}
\end{equation}%
where the effective optomechanical coupling $G_{j}$ is chosen, for
simplicity, to be $G_{j}=g_{j}\left\vert \langle \tilde{c}_{j}\rangle
\right\vert $.

The solution of the linearized quantum Langevin equations can be obtained as
$\mathcal{U}(t)=\mathcal{M}(t)\mathcal{U}(0)+\int_{0}^{t^{\prime }}dt%
\mathcal{M}(t)\mathcal{N}(t^{\prime }-t)$ where $\mathcal{M}(t)=\exp (%
\mathcal{K}t)$ \cite{vitali}$.$ The system is stable if and only if all real
parts of the eigenvalues of the matrix $\mathcal{K}$\ are negative, so that $%
\mathcal{M}(t\rightarrow \infty )=0$. The stability conditions can be
derived applying the Routh-Hurwitz criterion \cite{RH}, where details of the
calculations are given in Appendix A.

Since the dynamics of the system is linearized around the steady state, and
the operators $\widetilde{\hat{\zeta}}_{j}^{in}$ and $\digamma _{j}$ are
zero-mean quantum Gaussian noises, therefore the steady state of the quantum
fluctuations is a zero-mean four-mode Gaussian state described by its $%
8\times 8$ covariance matrix $\mathcal{\vartheta }$ of elements $\mathcal{%
\vartheta }_{ii^{\prime }}=\left( \langle \mathcal{U}_{i}(\infty )\mathcal{U}%
_{i^{\prime }}(\infty )+\mathcal{U}_{i^{\prime }}(\infty )\mathcal{U}%
_{i}(\infty )\rangle \right) /2$ \cite{vitali}. When the system is stable,
we obtain
\begin{equation}
\mathcal{\vartheta }_{ii^{\prime }}=\sum_{k,k^{\prime }}\int_{0}^{\infty
}ds\int_{0}^{\infty }ds_{^{ik}}^{\prime }\mathcal{M}(s)\mathcal{M}%
_{^{i^{\prime }k^{\prime }}}(s^{\prime })\Phi _{^{kk^{\prime }}}(s-s^{\prime
}),  \label{e17}
\end{equation}%
where $\Phi _{^{kk^{\prime }}}(s-s^{\prime })=(\langle \mathcal{N}_{k}(s)%
\mathcal{N}_{k^{\prime }}(s^{\prime })+\mathcal{N}_{k^{\prime }}(s^{\prime })%
\mathcal{N}_{k}(s)\rangle )/2=\mathcal{R}_{kk^{\prime }}\delta (s-s^{\prime
})$, with $\mathcal{R}_{kk^{\prime }}$ being the elements of the diffusion
matrix $\mathcal{R}$ \cite{vitali}. Using the correlation properties of the
operators $\widetilde{\hat{\zeta}}_{j}^{in}$ and $\digamma _{j}$, we show
that $\mathcal{R=R}_{m}\oplus \mathcal{R}_{c}$ with $\mathcal{R}_{m}=\left(
\gamma _{m_{1}}(2n_{\text{th},1}+1)\mbox{$1
\hspace{-1.0mm}  {\bf l}$}_{2}\oplus \gamma _{m_{2}}(2n_{\text{th},2}+1)%
\mbox{$1
\hspace{-1.0mm}  {\bf l}$}_{2}\right) $ and
\begin{equation}
\mathcal{R}_{c}=\left(
\begin{array}{cccc}
\kappa _{1}+\Xi _{11} & 0 & -\frac{\Xi _{12}+\Xi _{21}}{2} & 0 \\
0 & \kappa _{1}+\Xi _{11} & 0 & \frac{\Xi _{12}+\Xi _{21}}{2} \\
-\frac{\Xi _{12}+\Xi _{21}}{2} & 0 & \kappa _{2}+\Xi _{22} & 0 \\
0 & \frac{\Xi _{12}+\Xi _{21}}{2} & 0 & \kappa _{2}+\Xi _{22}%
\end{array}%
\right) .  \label{D-k}
\end{equation}

When the stability conditions are satisfied, Equation (\ref{e17}) becomes $%
\mathcal{\vartheta }=\int_{0}^{\infty }ds\mathcal{M}(s)\mathcal{RM}(s)^{%
\mathrm{T}}$ which is equivalent to the Lyapunov equation \cite{genes}
\begin{equation}
\mathcal{K\vartheta +\vartheta K}^{\mathrm{T}}=-\mathcal{R}.
\label{Lyapunov}
\end{equation}

The covariance matrix $\mathcal{\vartheta }$, solution of Equation (\ref%
{Lyapunov}), can be expressed as
\begin{equation}
\mathcal{\vartheta =}\left[ \mathcal{\vartheta }_{ij}\right] _{8\times 8}%
\mathcal{=}\left(
\begin{array}{cccc}
\mathcal{\vartheta }_{m_{1}} & \mathcal{\vartheta }_{m_{12}} & \mathcal{%
\vartheta }_{m_{1}c_{1}} & \mathcal{\vartheta }_{m_{1}c_{2}} \\
\mathcal{\vartheta }_{m_{12}}^{\mathrm{T}} & \mathcal{\vartheta }_{m_{2}} &
\mathcal{\vartheta }_{m_{2}c_{1}} & \mathcal{\vartheta }_{m_{2}c_{2}} \\
\mathcal{\vartheta }_{m_{1}c_{1}}^{\mathrm{T}} & \mathcal{\vartheta }%
_{m_{2}c_{1}}^{\mathrm{T}} & \mathcal{\vartheta }_{c_{1}} & \mathcal{%
\vartheta }_{c_{12}} \\
\mathcal{\vartheta }_{m_{1}c_{2}}^{\mathrm{T}} & \mathcal{\vartheta }%
_{m_{2}c_{2}}^{\mathrm{T}} & \mathcal{\vartheta }_{c_{12}}^{\mathrm{T}} &
\mathcal{\vartheta }_{c_{2}}%
\end{array}%
\right) ,  \label{CM}
\end{equation}%
where the $2\times 2$ blocks matrices $\mathcal{\vartheta }_{m_{1}}$ and $%
\mathcal{\vartheta }_{m_{2}}$ ($\mathcal{\vartheta }_{c_{1}}$ and $\mathcal{%
\vartheta }_{c_{2}}$) describe the first and second mechanical(cavity)
modes, while the correlations between them are described by the matrix $%
\mathcal{\vartheta }_{m_{12}}$($\mathcal{\vartheta }_{c_{12}}$). The matrix $%
\mathcal{\vartheta }_{m_{i}c_{j}}$ ($i,j\in \{1,2\}$) represents the
correlations between the $i\mathrm{th}$ mechanical mode and the $j\mathrm{th}
$ cavity mode.

As mentioned above, in this work we exploit the entanglement between the
two-mode laser emitted during the cascade transitions of the atomic system
to generate asymmetric Gaussian quantum steering between the two mechanical
modes $m_{1}$ and $m_{2}$, where their corresponding covariance matrix $%
\mathcal{\vartheta }_{m}$ can be deduced by tracing over the cavity modes
elements in Equation (\ref{CM}). Hence, we get
\begin{equation}
\mathcal{\vartheta }_{m}=\left(
\begin{array}{cc}
\mathcal{\vartheta }_{m_{1}} & \mathcal{\vartheta }_{m_{12}} \\
\mathcal{\vartheta }_{m_{12}}^{\mathrm{T}} & \mathcal{\vartheta }_{m_{2}}%
\end{array}%
\right) .  \label{MCM}
\end{equation}

\section{Gaussian Quantum Steering\label{SecIV}}

A bipartite state $\varrho _{AB}$, shared between two observers Alice and
Bob, is steerable from $A\rightarrow B$, i.e., Alice can steer Bob's states
via implementing a set of measurements $M_{A}$ on her side, if it is
impossible for every pair of local observables $r_{A}\in M_{A}$ on $A$ and $%
r_{B}$ (arbitrary) on $B$, with outcomes $r_{A}^{\prime }$ and $%
r_{B}^{\prime }$, respectively, to write the joint probability as $%
p(r_{A}^{\prime },r_{B}^{\prime }|r_{A},r_{B},\varrho _{AB})=\sum_{\alpha
}p_{\alpha }p(r_{A}^{\prime }|r_{A},\alpha )p(r_{B}^{\prime }|r_{B},\varrho
_{\alpha })$. In other words, at least one measurement pair $r_{A}$ and $%
r_{B}$ should violate this expression if $p_{\alpha }$ is fixed across all
measurements with $p_{\alpha }$ and $p(r_{A}^{\prime }|r_{A},\alpha )$ are
probability distributions and $p(r_{B}^{\prime }|r_{B},\varrho _{\alpha })$
represents the conditional probability distribution corresponding to the
extra condition of being evaluated on the state $\varrho _{\alpha }$. It has
been shown by Wiseman \textit{et al.} \cite{wiseman} that an arbitrary
bipartite Gaussian state $\varrho _{m_{1}m_{2}}$ with covariance matrix $%
\mathcal{\vartheta }_{m}$ is steerable under Gaussian measurements
implemented on party $m_{1}$ if and only if the expression $\mathcal{%
\vartheta }_{m}+\mathrm{i}(\Pi _{m_{1}}\oplus \Pi _{m_{2}})\geqslant \mathbf{%
0}$ is violated, with $\Pi _{m_{1}}=\left(
\begin{array}{cc}
0 & 0 \\
0 & 0%
\end{array}%
\right) $ and $\Pi _{m_{2}}$ $=\left(
\begin{array}{cc}
0 & 1 \\
-1 & 0%
\end{array}%
\right) $. On the basis of this constraint, Kogias \textit{et al}. \cite%
{kogias} have proposed a computable measure to quantify the amount by which
the state $\varrho _{m_{1}m_{2}}$ is steerable under Gaussian measurements
performed on party $m_{1}$, i.e.,
\begin{equation}
\mathcal{G}^{m_{1}\rightarrow m_{2}}:=\max \left[ 0,-\sum\limits_{j:\bar{\nu}%
_{j}^{m_{2}}<1}\ln \left\{ {\bar{\nu}_{j}^{m_{2}}}\right\} \right] ,
\label{e20}
\end{equation}%
with $\{\bar{\nu}_{j}^{m_{2}}\}$ being the symplectic spectra of the Schur
complement $\mathcal{\vartheta }_{m_{2}}-\mathcal{\vartheta }_{m_{12}}^{%
\mathrm{T}}\mathcal{\vartheta }_{m_{1}}^{-1}\mathcal{\vartheta }_{m_{12}}$
of party $m_{1}$ in the matrix $\mathcal{\vartheta }_{m}$ (\ref{MCM}).
Notice that $\mathcal{G}^{m_{1}\rightarrow m_{2}}$ is monotone under
Gaussian local operations and classical communication, and it vanishes if
the state $\varrho _{m_{1}m_{2}}$ is nonsteerable under Gaussian
measurements implemented on $m_{1}$. When $m_{1}$ and $m_{2}$ are single
modes, Equation (\ref{e20}) acquires the compact form $\mathcal{G}%
^{m_{1}\rightarrow m_{2}}=\max \left[ 0,\frac{1}{2}\ln \frac{\det \mathcal{%
\vartheta }_{m_{1}}}{4\det \mathcal{\vartheta }_{m}}\right] $ \cite{kogias},
where the steering from $m_{2}\rightarrow m_{1}$ can be obtained by changing
the roles of the modes $m_{1}$ and $m_{2}$ in Equation (\ref{e20}), i.e., $%
\mathcal{G}^{m_{2}\rightarrow m_{1}}=\max \left[ 0,\frac{1}{2}\ln \frac{\det
\mathcal{\vartheta }_{m_{2}}}{4\det \mathcal{\vartheta }_{m}}\right] $. A
nonzero value of $\mathcal{G}^{m_{i}\rightarrow m_{j}}$ means that the state
$\varrho _{m_{1}m_{2}}$ is steerable from $m_{i}\rightarrow m_{j}$ under
Gaussian measurements performed on mode $m_{i}$. Hence, three cases can be
distinguished: (\textit{1}) two-way steering, where the state $\varrho
_{m_{1}m_{2}}$ is steerable in both directions $m_{i(j)}\rightarrow m_{j(i)}$%
, i.e., $\mathcal{G}^{m_{i(j)}\rightarrow m_{j(i)}}>0$, (\textit{2}) no-way
steering, where $\mathcal{G}^{m_{i(j)}\rightarrow m_{j(i)}}=0$, and (\textit{%
3}) one-way steering, where the state $\varrho _{m_{1}m_{2}}$ is steerable
solely in one direction, that is, $\mathcal{G}^{m_{i}\rightarrow m_{j}}>0$
with $\mathcal{G}^{m_{j}\rightarrow m_{i}}=0$. We emphasize that pure
entangled states cannot in general display one-way steering behavior, since
they can always be transformed into a symmetric form through a local basis
change employing the Schmidt decomposition \cite{kogias}. Moreover, exploiting the method of homodyne detection \cite{vitali,solimeno}, the covariance matrix (\ref{MCM}) of the two mechanical modes $m_{1}$ and $m_{2}$ can be determined experimentally, which allows us to estimate numerically the steerabilities $\mathcal{G}^{m_{1(2)}\rightarrow m_{2(1)}}$.

Notice that varying the optical decay rates $\kappa _{1}$ and $\kappa _{2}$
is often employed as a main method to adjust the direction of one-way
steering \cite{QHe,WDeng,ZYang,SWu,HTan}. Such method, not only induces
additional losses and noises which affects the degree of the generated
steering, but also makes the experimental operations more complicated since
the optical losses are mainly related to the surface roughness, impurities
and defects of mirror materials \cite{aspelmayer}. In the scheme that we
consider here, we will show that the direction of one-way steering could be
controlled via the effective optomechanical coupling strengths $G_{1}$ and $%
G_{2}$ or the values of the means thermal occupations $n_{\text{th},1}$ and $%
n_{\text{th},2}$. The first method could be achieved via controlling the
lasers drive powers $\wp_{1} $ and $\wp _{2}$, while the second could be
achieved via adjusting the temperatures $T_{1}$ and $T_{2} $ of the
mechanical baths, which is more practicable in experimental operations.

\begin{figure}[t]
\centerline{\includegraphics[width=0.4\columnwidth,height=4cm]{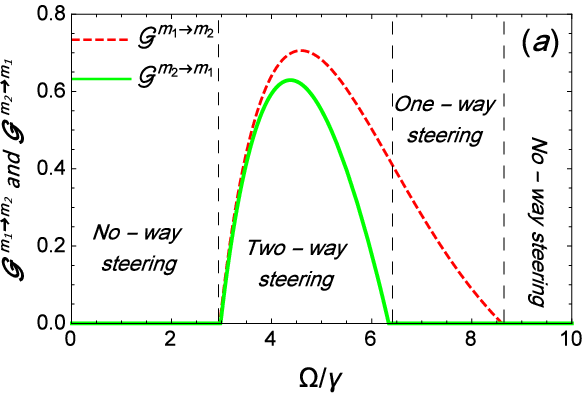}
\includegraphics[width=0.4\columnwidth,height=4cm]{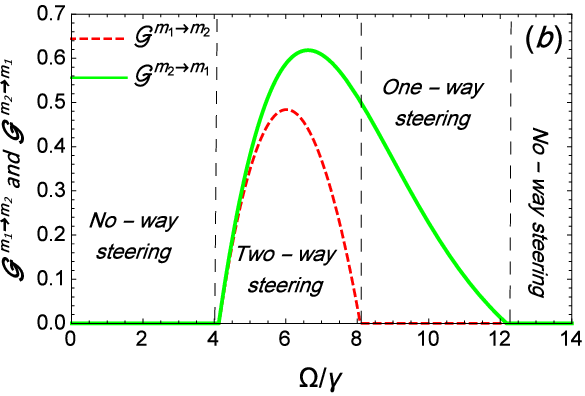}}
\caption{Plot of the Gaussian steerabilities $\mathcal{G}^{m_{1}\rightarrow
m_{2}}$ (red dashed curve) and $\mathcal{G}^{m_{2}\rightarrow m_{1}}$ (green
solid curve) versus the normalized driving field strength $\Omega /\protect%
\gamma $ using $\protect\kappa =2\protect\pi \times 215~\mathrm{kHz}$, $%
\protect\gamma _{m}=2\protect\pi \times 140~\mathrm{Hz}$, $\protect\omega %
_{m}=2\protect\pi \times 947~\mathrm{kHz}$, $G_{1,2}/\protect\omega %
_{m}=0.25 $, $\gamma=1.7~\text{MHz}$, and $\mathcal{A}=250~\mathrm{MHz}$. We used $n_{\text{th},1}=15$ which corresponds to $T_{1}=0.7~\textrm{mK}$ and $n_{\text{th},2}=5$ which corresponds to $T_{2}=0.3~\textrm{mK}$ in (a), while we used $n_{\text{th},1}=5$ and $n_{\text{th}%
,2}=15$ in (b). The stronger two-way steering ($\mathcal{G}%
^{m_{1(2)}\rightarrow m_{2(1)}}>0$) observed over a wide range of $\Omega /%
\protect\gamma $ in panels (a) and (b) is demonstrated to be a fundamental
resource required for teleporting a coherent state with fidelity beyond the
nocloning threshold \protect\cite{QHe}. While, the different degree of
steering in both directions $m_{1}\rightleftarrows m_{2}$, including one-way
steering behavior, is proved to provide the asymmetric guaranteed key rate
achievable in a practical 1SDI-QKD \protect\cite{kogias}. We should mention that the temperatures $T_{1}$ and $T_{2}$ employed in this figure are around $1~\textrm{mK}$, which has been achieved experimentally in \cite{Norte} with mechanical frequency very close to that used in the present work.}
\label{fig2}
\end{figure}

For numerical estimation of the steerabilities $\mathcal{G}%
^{m_{1}\rightarrow m_{2}}$ and $\mathcal{G}^{m_{2}\rightarrow m_{1}}$, we
have taken identical mirrors with mass $\mu _{1,2}=145~\mathrm{ng}$,
frequency $\omega _{m_{1,2}}=2\pi \times 947~\mathrm{kHz}$, and damping rate
$\gamma _{m_{1,2}}=2\pi \times 140~\mathrm{Hz}$. The cavities, having
identical decay rates $\kappa _{1,2}=2\pi \times 215~\mathrm{kHz}$,
equilibrium lengths $\ell _{1}=0.532~\mathrm{mm}$ and $\ell _{2}=0.405~%
\mathrm{mm}$, frequencies $\nu _{1}=2\pi \times 4.32\times 10^{14}~\mathrm{Hz%
}$ and $\nu _{2}=2\pi \times 4.33\times 10^{14}~\mathrm{Hz} $, are pumped by
input lasers of frequencies $\omega _{L_{1}}=2\pi \times 3.7\times 10^{14}~%
\mathrm{Hz}$ and $\omega _{L_{2}}=2\pi \times 2.82\times 10^{14}~\mathrm{Hz}$
\cite{groblacher,cohadon}. The atomic decay rates $\gamma _{1,2}=1.7~\text{%
MHz}$, the atom--field coupling strengths $\varsigma _{1,2}=15~\text{MHz}$,
the rate at which the atoms are injected into the cavity $r_{0}=1.6~\mathrm{%
MHz}$, so that the linear gain coefficient $\mathcal{A}=2r_{0}\varsigma
^{2}/\gamma ^{2}\approx 250~\mathrm{MHz}$ \cite{nha}.

\begin{figure}[h]
\centerline{\includegraphics[width=0.4\columnwidth,height=4cm]{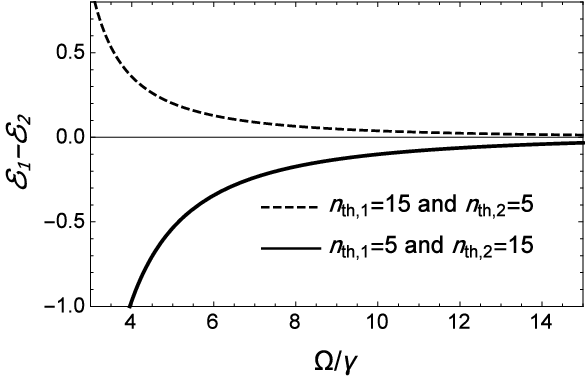}}
\caption{Plot of the difference $\mathcal{E}_{1}-\mathcal{E}_{2}$, in units
of $\frac{1}{2}\hbar \protect\omega _{m}$, where $\mathcal{E}_{j}=\frac{%
\hbar \protect\omega _{m}}{2}(\langle \protect\delta \tilde{q}%
_{m_{j}}^{2}\rangle +\langle \protect\delta \tilde{p}_{m_{j}}^{2}\rangle)$
is the energy of the $j\text{th}$ mechanical mode. The black dashed(solid)
curve is plotted under the same conditions as in Figure \protect\ref{fig2}%
(a)(Figure \protect\ref{fig2}(b)).}
\label{fig3}
\end{figure}

In Figure \ref{fig2} we plot the effect of the normalized driving field
strength $\Omega /\gamma $ on the steerabilities $\mathcal{G}%
^{m_{1}\rightarrow m_{2}}$ and $\mathcal{G}^{m_{2}\rightarrow m_{1}}$ using $%
G_{1,2}/\omega _{m}=0.25$ for the effective optomechanical coupling
strengths. For the thermal occupancies, we used $n_{\text{th},1}=15$ which corresponds to $T_{1}=0.7~\textrm{mK}$ and $n_{%
\text{th},2}=5$ which corresponds to $T_{2}=0.3~\textrm{mK}$ in Figure \ref{fig2}(a), while we used $n_{\text{th},1}=5$ and $n_{%
\text{th},2}=15$ in Figure \ref{fig2}(b). Even though there is no direct
interaction between the mechanical modes $m_{1}$ and $m_{2}$, Gaussian
quantum steering could be detected between them due to quantum coherence
transfer from the three-level laser to the state $\varrho _{m_{1}m_{2}}$ via
the optomechanical coupling. Manifestly, there exists a minimum strength of
the driving field, and then a minimum of atomic coherence for which the
state $\varrho _{m_{1}m_{2}}$ is steerable. Furthermore, by increasing
gradually the drive amplitude $\Omega $---from the required minimum---the
steerabilities $\mathcal{G}^{m_{1}\rightarrow m_{2}}$ and $\mathcal{G}%
^{m_{2}\rightarrow m_{1}}$ arise and undergo a resonance-like behavior. This
indicates that the steering between the two mirrors can be controlled with
the driven filed that couples the atomic coherence to the cavity modes.
Essentially, Figure \ref{fig2} shows that inferring about the mechanical
mode $m_{i}$ based on measurements performed on mode $m_{j}$ is fully
different from the reverse operation, i.e., $\mathcal{G}^{m_{1}\rightarrow
m_{2}}\neq \mathcal{G}^{m_{2}\rightarrow m_{1}}$, where two-way steering
(i.e., $\mathcal{G}^{m_{1(2)}\rightarrow m_{2(1)}}>0$) and one-way steering
(i.e., $\mathcal{G}^{m_{i}\rightarrow m_{j}}>0$ and $\mathcal{G}%
^{m_{j}\rightarrow m_{i}}=0$ for $i\neq j=1,2$) could be observed. From an
operational point of view, one-way steering observed from $m_{i}\rightarrow
m_{j}$ implies that the observer owning mode $m_{i}$ and that owning mode $%
m_{j}$ can perform identical Gaussian measurements on their own modes, however get conflicting outcomes. Moreover, the
former can convince the latter that the state $\varrho _{m_{1}m_{2}}$ is
entangled, while the converse is not true. This is partly due to the
asymmetric form of the matrix $\mathcal{\vartheta }$ (\ref{MCM}), where $%
\det \mathcal{\vartheta }_{m_{1}}\neq \det \mathcal{\vartheta }_{m_{2}}$,
and partly due to the definition of the aspect of quantum steering in terms
of the EPR paradox \cite{wiseman,kogias}. We notice that asymmetric steering
is proved to supply security in 1SDI-QKD protocol, where the measurement
apparatus of one device is untrusted \cite{kogias}.

\begin{figure}[th]
\centerline{\includegraphics[width=0.4\columnwidth,height=4cm]{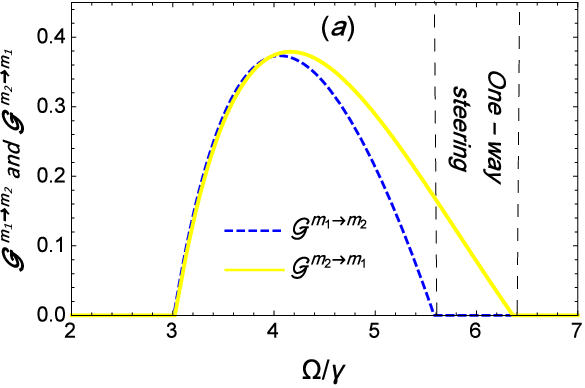}
		\includegraphics[width=0.4\columnwidth,height=4cm]{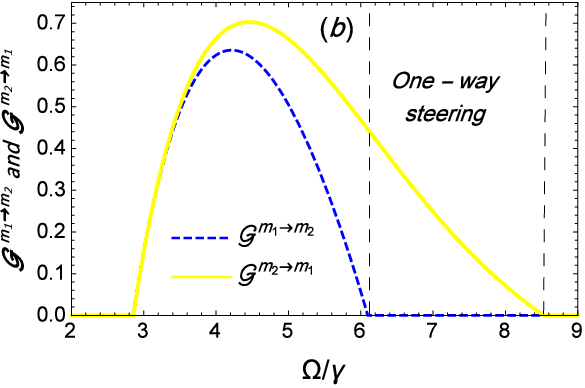}}
\caption{Plot of the Gaussian steerabilities $\mathcal{G}^{m_{1}\rightarrow
m_{2}}$ (blue dashed curve) and $\mathcal{G}^{m_{2}\rightarrow m_{1}}$
(yellow solid curve) versus the normalized driving field strength $\Omega /%
\protect\gamma $ using $\protect\kappa =2\protect\pi \times 215~\mathrm{kHz}$%
, $\protect\gamma _{m}=2\protect\pi \times 140~\mathrm{Hz}$, $\protect\omega %
_{m}=2\protect\pi \times 947~\mathrm{kHz}$, $\mathcal{A}=250~\mathrm{MHz}$,
and $n_{\text{th},1,2}=15$. For the effective optomechanical coupling
strengths, we used $G_{1,2}/\protect\omega _{m}=G/\protect\omega _{m}=0.25$
in (a) and $G/\protect\omega _{m}=0.35$ in (b). This figure clearly shows
the advantage of the mediation of the effective optomechanical coupling
strength in observing one-way steering through a wide range of $\Omega /%
\protect\gamma $, which allows us the implementation of more one-way quantum
information tasks.}
\label{fig4}
\end{figure}

Strikingly, in Figure \ref{fig2}(a) where we used $n_{\text{th},1}=15$ and $%
n_{\text{th},2}=5$, the steering $\mathcal{G}^{m_{1}\rightarrow m_{2}}$
remains greater than $\mathcal{G}^{m_{2}\rightarrow m_{1}}$ and one-way
steering is occurred from $m_{1}\rightarrow m_{2}$. While, $\mathcal{G}%
^{m_{1}\rightarrow m_{2}}$ remains less than $\mathcal{G}^{m_{2}\rightarrow
m_{1}}$ and one-way steering is occurred in the reverse direction $%
m_{2}\rightarrow m_{1}$ in Figure \ref{fig2}(b) where we used $n_{\text{th}%
,1}=5$ and $n_{\text{th},2}=15$. Therefore, an appropriate choice of the
temperatures $T_{1}$ and $T_{2}$ of the mechanical baths may provide a
flexible and feasible experimental manner to manipulate the direction of
one-way steering. To better explain the direction of one-way steering
observed in Figure \ref{fig2}, we analyse the sign of the difference $%
\mathcal{E}_{1}-\mathcal{E}_{2}$ where $\mathcal{E}_{j}=\frac{\hbar \omega
_{m}}{2}(\langle \delta \tilde{q}_{m_{j}}^{2}\rangle +\langle \delta \tilde{p%
}_{m_{j}}^{2}\rangle )$ is the mean energy of the $j$\textrm{th} mechanical
mode. In Figure \ref{fig3} we plot $\mathcal{E}_{1}-\mathcal{E}_{2}$, in
units of $\hbar \omega _{m}/2$, under the same conditions as in Figure \ref%
{fig2}. As can be seen, the direction of one-way steering is strongly
influenced by the sign of the difference $\mathcal{E}_{1}-\mathcal{E}_{2}$.
Indeed, in the situation where $\mathcal{E}_{1}-\mathcal{E}_{2}$ remains
positive, which corresponds to $n_{\text{th},1}=15$ and $n_{\text{th},2}=5$,
one-way steering is occurred from $m_{1}\rightarrow m_{2}$ in Figure \ref%
{fig2}(a). While for $n_{\text{th},1}=5$ and $n_{\text{th},2}=15$, $\mathcal{%
E}_{1}-\mathcal{E}_{2}$ remains negative and one-way steering is occurred
from $m_{2}\rightarrow m_{1}$ in Figure \ref{fig2}(b). Physically this could
be interpreted as follows: for a two-mode Gaussian state subject to Gaussian
measurements, the mode having a higher fluctuation is more difficult to be
steered by the other mode having a lower fluctuation.

Next, in Figure \ref{fig4} we plot the steerabilities $\mathcal{G}%
^{m_{1}\rightarrow m_{2}}$ and $\mathcal{G}^{m_{2}\rightarrow m_{1}}$ versus
$\Omega /\gamma $ using $n_{\text{th},1,2}=15$ and two different values of
the common effective optomechanical coupling $G_{1,2}/\omega _{m}\equiv
G/\omega _{m}$, namely, $G/\omega _{m}=0.25$ in Figure \ref{fig4}(a) and $%
G/\omega _{m}=0.35$ in Figure \ref{fig4}(b). In contrary to the results
depicted in Figure \ref{fig2}, Figure \ref{fig4} shows, under the use of
balanced thermal occupations and balanced effective optomechanical
couplings, the occurrence of one-way steering only from $m_{2}\rightarrow
m_{1}$. Importantly, the increasing of $G/\omega _{m}$ from $0.25$ to $0.35$
enlarges drastically the range of $\Omega /\gamma $ corresponding to one-way
steering. This indicates the advantage of the mediation of the
optomechanical coupling in achieving one-way steering through a wide range
of $\Omega /\gamma $, which provides the asymmetric guaranteed key rate
achievable within a practical 1SDI-QKD \cite{kogias}.

\begin{figure}[th]
\centerline{\includegraphics[width=0.4\columnwidth,height=4cm]{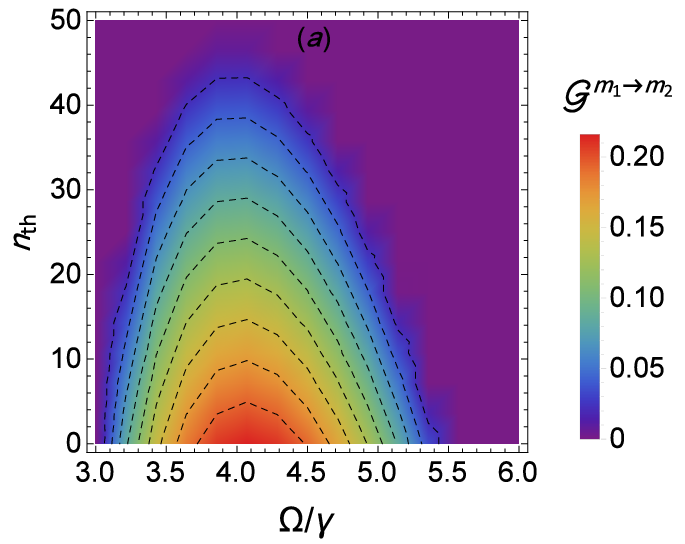}
		\includegraphics[width=0.4\columnwidth,height=4cm]{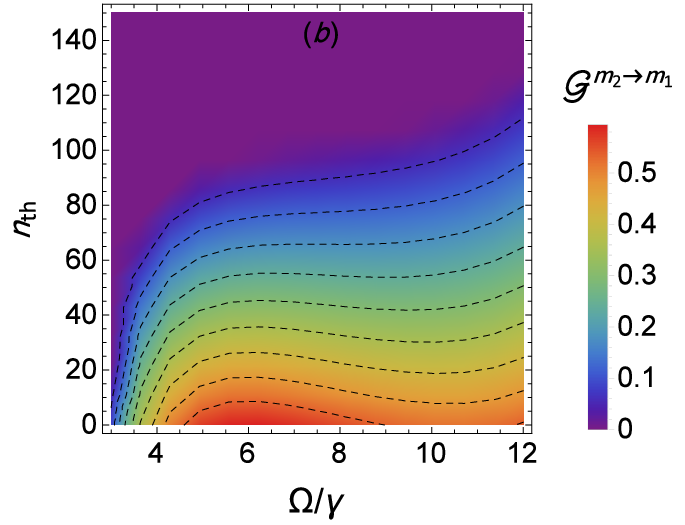}}
\centerline{\includegraphics[width=0.4\columnwidth,height=4cm]{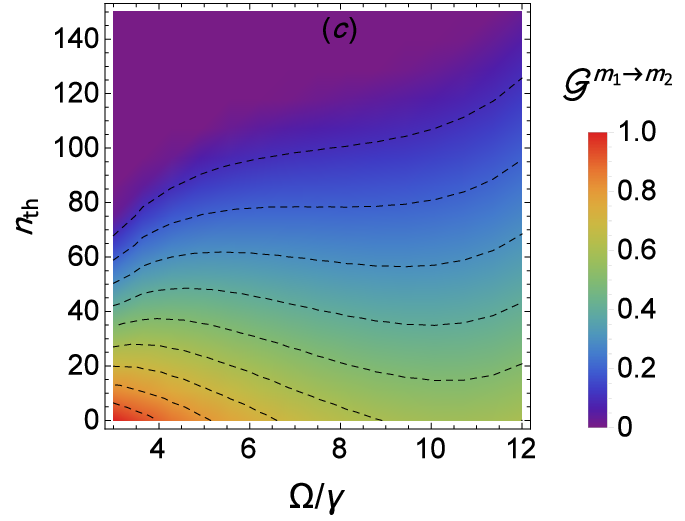}
		\includegraphics[width=0.4\columnwidth,height=4cm]{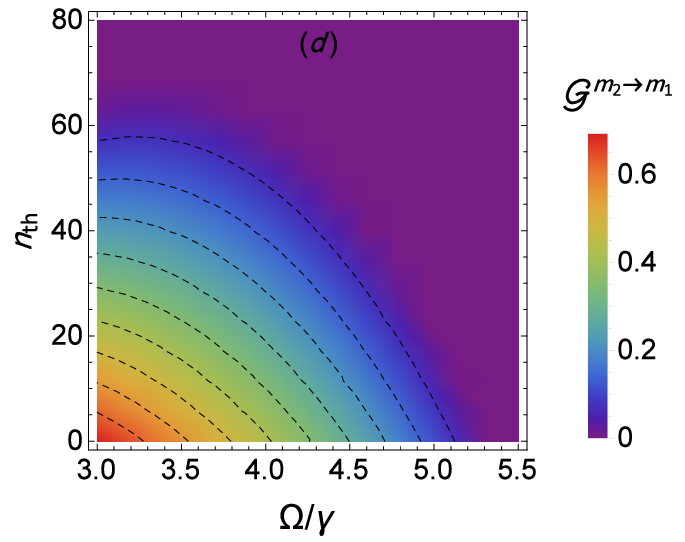}}
\caption{ Density plot of the Gaussian steerabilities $\mathcal{G}%
^{m_{1}\rightarrow m_{2}}$ and $\mathcal{G}^{m_{2}\rightarrow m_{1}}$ versus
the normalized driving field strength $\Omega /\protect\gamma $ and the
common mean thermal phonon number $n_{\text{th},1,2}=n_{\text{th}}$ using $%
\protect\kappa =2\protect\pi \times 215~\mathrm{kHz}$, $\protect\gamma _{m}=2%
\protect\pi \times 140~\mathrm{Hz}$, $\protect\omega _{m}=2\protect\pi %
\times 947~\mathrm{kHz}$, and $\mathcal{A}=250~\mathrm{MHz}$. For the
effective optomechanical coupling strengths, we used $G_{1}/\protect\omega %
_{m}=0.25$ and $G_{2}/\protect\omega _{m}=0.35$ in [(a) and (b)], while $%
G_{1}/\protect\omega _{m}=0.35$ and $G_{2}/\protect\omega _{m}=0.25$ in [(c)
and (d)]. In the appendix, we show that the chosen values of $G_{1}$, $G_{2}$%
, and $\Omega /\protect\gamma $ fulfill the stability conditions derived
using the Routh-Hurwitz criterion \protect\cite{RH}.}
\label{fig5}
\end{figure}

It is well known that real quantum systems are inevitably affected by their
surrounding environments which leads to the degradation of quantum
correlations via decoherence process. This is a challenging issue for
creating and preserving quantum correlations in open systems which is of
great importance for quantum information processing. In Figure \ref{fig5},
we study $\mathcal{G}^{m_{1}\rightarrow m_{2}}$ and $\mathcal{G}%
^{m_{2}\rightarrow m_{1}}$ under influence of the common mean thermal phonon
number $n_{\mathrm{th,1,2}}=n_{\mathrm{th}}$ and the normalized driving
field strength $\Omega /\gamma $ using $G_{1}/\omega _{m}=0.25$ and $%
G_{2}/\omega _{m}=0.35$ in [Figures \ref{fig5}(a) and \ref{fig5}(b)], while $%
G_{1}/\omega _{m}=0.35$ and $G_{2}/\omega _{m}=0.25$ in [Figures \ref{fig5}%
(c) and \ref{fig5}(d)]. It is vividly shown that both steerabilities $%
\mathcal{G}^{m_{1}\rightarrow m_{2}}$ and $\mathcal{G}^{m_{2}\rightarrow
m_{1}}$ are maximum close to $n_{\mathrm{th}}=0$, however with increasing $%
n_{\mathrm{th}}$, they decrease gradually and eventually disappear.

Interestingly, Figure \ref{fig5} shows that $\mathcal{G}^{m_{1}\rightarrow
m_{2}}$ and $\mathcal{G}^{m_{2}\rightarrow m_{1}}$ exhibit rather different
trends against thermal noise that, not only reduces the degree of $\mathcal{G%
}^{m_{1}\rightarrow m_{2}}$ and $\mathcal{G}^{m_{2}\rightarrow m_{1}}$, but
also it could play a constructive role in inducing one-way steering via a
suitable choice of the ratio $G_{1}/G_{2}$. Indeed, in [Figures \ref{fig5}%
(a) and \ref{fig5}(b)] where $G_{1}/G_{2}<1$, $\mathcal{G}^{m_{1}\rightarrow
m_{2}}$ vanishes entirely for $n_{\mathrm{th}}>45$, while $\mathcal{G}%
^{m_{2}\rightarrow m_{1}}$ still persists and it is significantly nonzero
for thermal occupancy up to $n_{\mathrm{th}}\approx 120$ meaning that the
state $\varrho _{m_{1}m_{2}}$ is one-way steerable from $m_{2}\rightarrow
m_{1}$ for $45\leqslant n_{\text{th}}\leqslant 120$. Whereas, [Figures \ref%
{fig5}(c) and \ref{fig5}(d)] show that for $G_{1}/G_{2}>1$, $\mathcal{G}%
^{m_{2}\rightarrow m_{1}}$ vanishes for $n_{\mathrm{th}}>60$, while $%
\mathcal{G}^{m_{1}\rightarrow m_{2}}$ is fairly robust against thermal noise
up to $n_{\mathrm{th}}\approx 145$ meaning that the state $\varrho
_{m_{1}m_{2}}$ is one-way steerable from $m_{1}\rightarrow m_{2}$ for $%
60\leqslant n_{\text{th}}\leqslant 145$. Then, one concludes that the
direction of one-way steering between the two mechanical modes $m_{1}$ and $%
m_{2}$ can be manipulated with the temperatures of the mechanical baths and
the optomechanical coupling strengths.

\begin{figure}[h]
\centerline{\includegraphics[width=0.4\columnwidth,height=4cm]{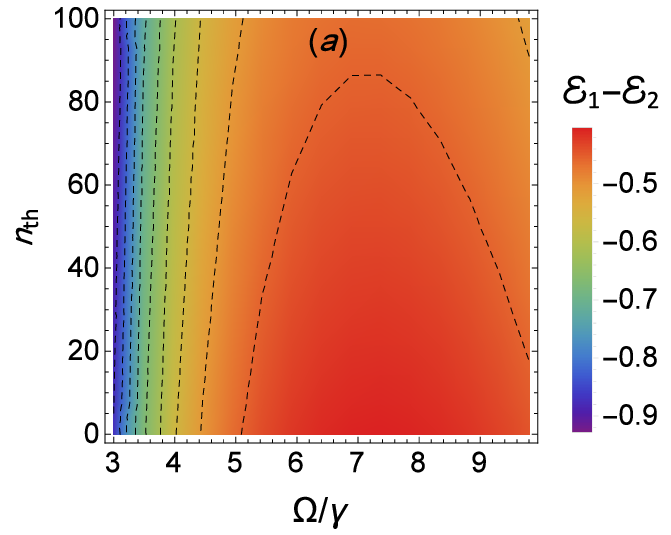}
\includegraphics[width=0.4\columnwidth,height=4cm]{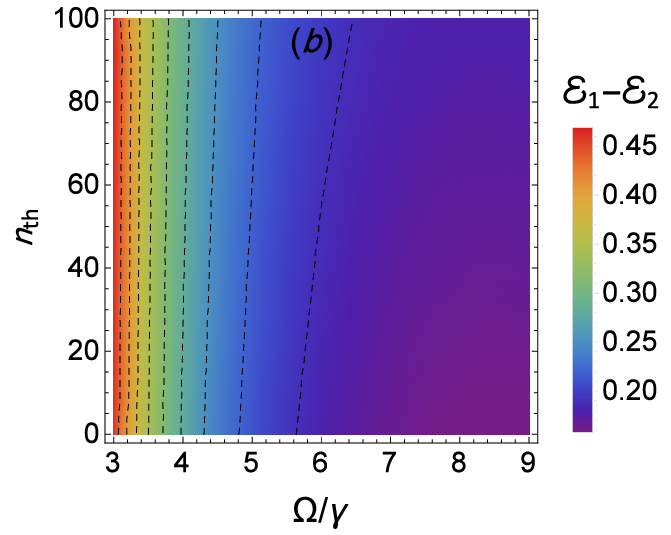}}
\caption{Plot of $\mathcal{E}_{1}-\mathcal{E}_{2}$, in units of $\frac{1}{2}%
\hbar\protect\omega_{m}$, versus $\Omega/\protect\gamma$ and $n_{\text{th}}$%
. The parameters in (a) and (b) are the same as in [Figures \protect\ref%
{fig5}(a) and \protect\ref{fig5}(b)] and [Figures \protect\ref{fig5}(c) and
\protect\ref{fig5}(d)], respectively.}
\label{fig6}
\end{figure}

In Figure \ref{fig6}(a) we plot the difference $\mathcal{E}_{1}-\mathcal{E}%
_{2}$ for the same conditions as in [Figures \ref{fig5}(a) and \ref{fig5}%
(b)], whereas in Figure \ref{fig6}(b) we plot $\mathcal{E}_{1}-\mathcal{E}%
_{2}$ for the same conditions as in [Figures \ref{fig5}(c) and \ref{fig5}%
(d)]. We also remark that the direction of one-way steering depends on the
sign of the difference $\mathcal{E}_{1}-\mathcal{E}_{2}$, i.e., in Figure %
\ref{fig6}(a) where $\mathcal{E}_{1}-\mathcal{E}_{2}<0$, one-way steering is
occurred from $m_{2}\rightarrow m_{1}$ in [Figures \ref{fig5}(a) and \ref%
{fig5}(b)], while in Figure \ref{fig6}(b) where $\mathcal{E}_{1}-\mathcal{E}%
_{2}>0$, one-way steering is occurred from $m_{1}\rightarrow m_{2}$ in
[Figures \ref{fig5}(c) and \ref{fig5}(d)]. This is in concordance with the
results obtained in Figures \ref{fig2} and \ref{fig3}.

Since the security of 1SDI-QKD protocol depends crucially on the direction
of steering, the presented scheme may be useful for such protocol, where
two-way steering can be regarded as part of a legitimate step of the
protocol \cite{branciard}, while one-way steering can be considered as an
attack from an adversarial party \cite{branciard}. On the other hand, the
change in the direction of steering may change the role played by two
persons during their communication employing 1SDI-QKD protocol, which is of
great practical importance in the security of such protocol \cite{branciard}.

With the stability conditions given in Appendix A and experimentally
accessible parameters \cite{groblacher,cohadon,QED1,QED2}, macroscopic asymmetric
Gaussian steering for two vibrating mirrors may be realized with the current
state-of-the-art experimental apparatus. Furthermore, by exploiting the
measurement strategy proposed in \cite{palomaki,vitali}, the covariance
matrix (\ref{MCM}) can be fully reconstructed using standard homodyne
detection \cite{solimeno}, which allows us to estimate numerically the
steerabilities $\mathcal{G}^{m_{1}\rightarrow m_{2}}$ and $\mathcal{G}%
^{m_{2}\rightarrow m_{1}}$.

\section{Conclusion}

\label{SecV}

We have investigated a scheme for generating asymmetric steering between two
non-interacting mechanical modes $m_{1}$ and $m_{2}$ by transferring quantum
correlations from a correlated-emission laser through radiation pressure.
The laser's gain medium consists of a set of nondegenerate three-level atoms
in a cascade configuration. By applying the master equation for a two-mode
laser, we derived the linearized quantum Langevin equations that describe
the optomechanical coupling between modes $m_{1}$ and $m_{2}$ and the two
cavity modes. These equations were then used to obtain the covariance matrix
for modes $m_{1}$ and $m_{2}$ in the steady state. Using realistic
experimental parameters, we demonstrated that both two-way and one-way
steering can be observed by varying the strength of the external field
driving the gain medium. Unlike most schemes that rely on optical losses to
control the direction of one-way steering, our findings indicate that the
direction of one-way steering can be adjusted on demand by modifying either
the optomechanical coupling strengths or the temperatures of the mechanical
baths, providing a flexible and practical approach for experimental
implementation. Additionally, we revealed that the direction of one-way
steering is influenced by mode fluctuations, with the mechanical mode
exhibiting higher fluctuations dominating the directionality.

Our results demonstrate a viable method for manipulating the direction of
one-way steering in a doubly resonant optomechanical cavity, which may have
potential applications in one-way quantum information tasks, such as one-way
quantum computation and communication.

\section*{Appendix A: Stability Analysis}

We provide here the stability analysis via the Routh--Hurwitz criterion \cite%
{RH} to ensure the choice of parameters made during the study of the
steerabilities $\mathcal{G}^{m_{1}\rightarrow m_{2}}$ and $\mathcal{G}%
^{m_{2}\rightarrow m_{1}}$. The Routh-Hurwitz criterion asserts that the
system can reach a stable steady state when all real parts of the
eigenvalues ${\chi}$ of the matrix $\mathcal{K}$ (\ref{e13}) are negative.
The eigenvalues ${\chi}$ are given by the roots of the characteristic
equation $\det (\mathcal{K}-\chi \mbox{$1
\hspace{-1.0mm}  {\bf l}$}_{8})=0$ which can be reduced to $a_{0}\chi
^{8}+a_{1}\chi ^{7}+a_{2}\chi ^{6}+a_{3}\chi ^{5}+a_{4}\chi ^{4}+a_{5}\chi
^{3}+a_{6}\chi ^{2}+a_{7}\chi +a_{8}=0$, where the coefficients $a_{i}$ are
too cumbersome to be reported here.

In fact, the Routh-Hurwitz criterion constrains the coefficients $%
a_{i}(i=0,..,8)$, and then the system parameters, through the Hurwitz
determinants obtained from the determinant
\begin{equation}
\Lambda _{n}=\left\vert
\begin{array}{cccccc}
a_{1} & a_{3} & a_{5} & \ldots & 0 & 0 \\
a_{0} & a_{2} & a_{4} & \ldots & 0 & \vdots \\
0 & a_{1} & a_{3} & \ldots & 0 & \vdots \\
0 & a_{0} & a_{2} & \ddots & a_{n} & \vdots \\
0 & \ldots & \ldots & a_{n-3} & a_{n-1} & 0 \\
0 & \ldots & \ldots & a_{n-4} & a_{n-2} & a_{n}%
\end{array}%
\right\vert ,  \label{Routh}
\end{equation}%
with $a_{0}=1$, $\Lambda _{1}=a_{1}$, $\Lambda _{2}=\left\vert
\begin{array}{cc}
a_{1} & a_{3} \\
a_{0} & a_{2}%
\end{array}%
\right\vert $, $\Lambda _{3}=\left\vert
\begin{array}{ccc}
a_{1} & a_{3} & a_{5} \\
a_{0} & a_{2} & a_{4} \\
0 & a_{1} & a_{3}%
\end{array}%
\right\vert $, ..., $\Lambda _{8}=\left\vert
\begin{array}{cccccccc}
a_{1} & a_{3} & a_{5} & a_{7} & 0 & 0 & 0 & 0 \\
a_{0} & a_{2} & a_{4} & a_{6} & a_{8} & 0 & 0 & 0 \\
0 & a_{1} & a_{3} & a_{5} & a_{7} & 0 & 0 & 0 \\
0 & a_{0} & a_{2} & a_{4} & a_{6} & a_{8} & 0 & 0 \\
0 & 0 & a_{1} & a_{3} & a_{5} & a_{7} & 0 & 0 \\
0 & 0 & a_{0} & a_{2} & a_{4} & a_{6} & a_{8} & 0 \\
0 & 0 & 0 & a_{1} & a_{3} & a_{5} & a_{7} & 0 \\
0 & 0 & 0 & a_{0} & a_{2} & a_{4} & a_{6} & a_{8}%
\end{array}%
\right\vert $.

\begin{figure}[th]
\centerline{\includegraphics[width=0.25\columnwidth,height=4cm]{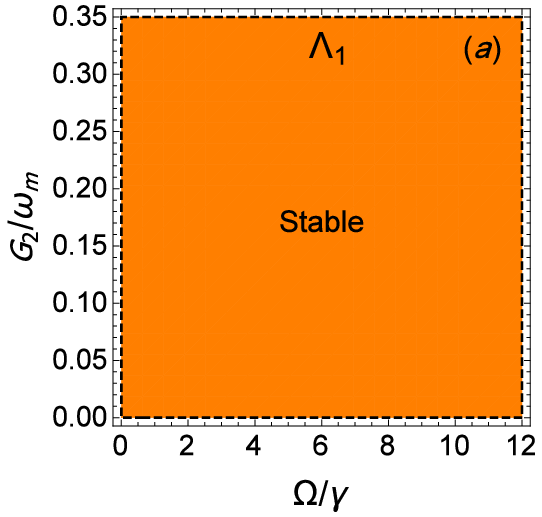}
		\includegraphics[width=0.25\columnwidth,height=4cm]{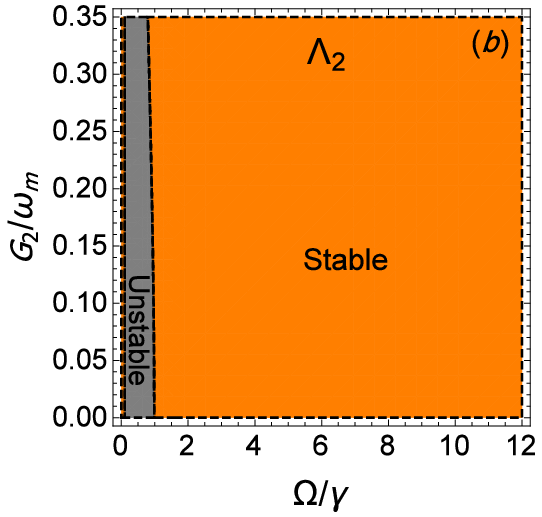}
\includegraphics[width=0.25\columnwidth,height=4cm]{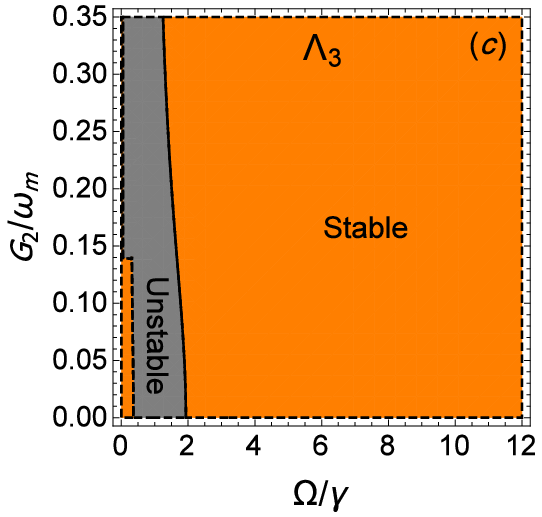}
\includegraphics[width=0.25\columnwidth,height=4cm]{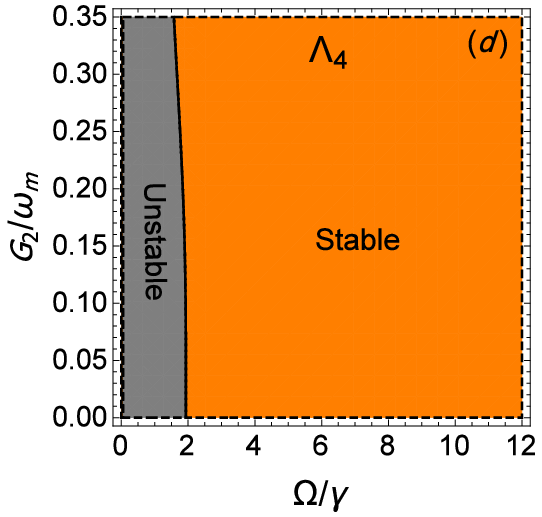}}
\centerline{\includegraphics[width=0.25\columnwidth,height=4cm]{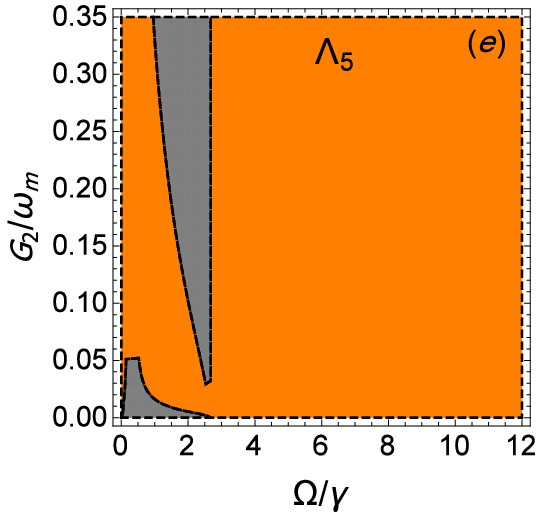}
		\includegraphics[width=0.25\columnwidth,height=4cm]{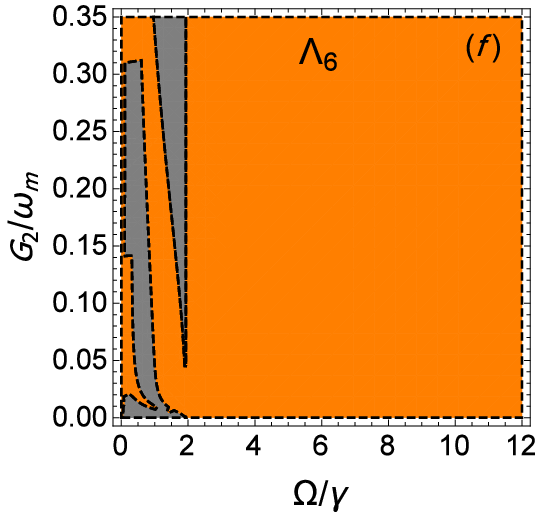}
\includegraphics[width=0.25\columnwidth,height=4cm]{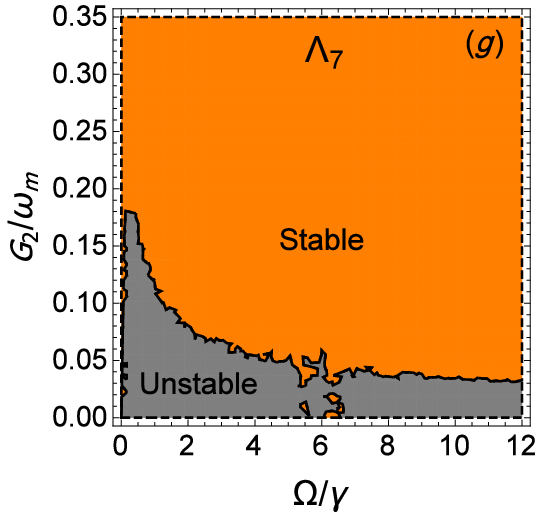}
\includegraphics[width=0.25\columnwidth,height=4cm]{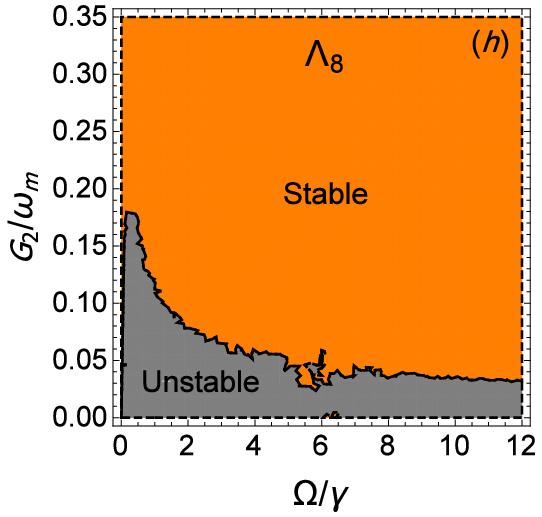}}
\caption{Plot of the Hurwitz determinants $\Lambda _{n}$ for $n=1,..,8$
versus the $\Omega /\protect\gamma $ and $G_{2}/\protect\omega _{m}$ using $%
G_{1}/\protect\omega _{m}=0.25$, $\protect\omega _{m}=2\protect\pi \times
947~\mathrm{kHz}$, $\protect\kappa =2\protect\pi \times 215~\mathrm{kHz}$, $%
\protect\gamma _{m}=2\protect\pi \times 140$, and $\mathcal{A}=250~\mathrm{%
MHz}$. The orange part is the stable area, while the gray one is the
unstable area.}
\label{fig7}
\end{figure}

In Figure \ref{fig7}, we plot the determinants $\Lambda _{n}$ for $n=1,..,8$
using $G_{1}/\omega _{m}=0.25$, $0\leqslant G_{2}/\omega _{m}\leqslant 0.35$%
, $0\leqslant \Omega /\gamma \leqslant 12$, $\omega _{m}=2\pi \times 947~%
\mathrm{kHz}$, $\kappa =2\pi \times 215~\mathrm{kHz}$, $\gamma _{m}=2\pi
\times 140~\mathrm{Hz}$, and $\mathcal{A}=250~\mathrm{MHz.}$ As can be seen
from in Figure \ref{fig7}(a), the determinant $\Lambda _{1}$ remains
positive, then according to the Routh-Hurwitz criterion, the real parts of
the eigenvalues of the matrix $\mathcal{K}$ (\ref{e13}) are all negative if
and only if the sequence of the determinants $\Lambda _{n}$ for $n=2$,.., $8$
are all positive. For $0.2\leqslant G_{2}/\omega _{m}\leqslant 0.35$ and $%
3\leqslant \Omega /\gamma \leqslant 12$, we have $\Lambda _{n}>0$ for $%
n=1,..,8$, meaning that the values of $G_{1}/\omega _{m}$, $G_{2}/\omega
_{m} $, and $\Omega /\gamma $, considered in our numerical simulations,
guarantee the stability of our system.

\section*{Appendix B: Derivation of Equation (\ref{hn})}

In this Appendix, employing the procedure outlined in refs. \cite{Louissel,Sargent}, we derive the master
equation of the reduced density matrix $\rho _{o_{1}o_{2}}$ for two-mode
laser $o_{1}$ and $o_{2}$ generated when a single three-level atom makes
transitions from $|\ell _{1}\rangle \rightarrow |\ell _{2}\rangle $ and $%
|\ell _{2}\rangle \rightarrow |\ell _{3}\rangle $, respectively [Figure \ref%
{fig1}(b)]. First, we adopt $\rho _{\left( \mathrm{sa,}o_{1}o_{2}\right)
}(t,t_{j})$ to represent the density matrix at time $t$ for the modes $o_{1}$
and $o_{2}$ plus a single atom pumped into the cavity at an earlier time $%
t_{j}$. Then, the density matrix of the atomic system plus the modes $o_{1}$
and $o_{2}$ in the cavity at time $t$ writes $\rho _{\left( \mathrm{sa,}%
o_{1}o_{2}\right) }(t)=r_{0}\sum\limits_{j}\rho _{\left( \mathrm{sa,}%
o_{1}o_{2}\right) }(t,t_{j})\Delta t,$ where $r_{0}\Delta t$ yields the
number of atoms injected into the cavity within a brief time interval $%
\Delta t$. The atoms are injected into the cavity at a rate $r_{0}$ and
removed after a time $\tau $, supposed, longer than the spontaneous emission
time \cite{sz}. During this time interval $\tau $, a single atom resonantly
interacts with the cavity modes of frequencies $\nu _{1}$ and $\nu _{2}$.

When $\Delta t\rightarrow 0$, the summation can be approximated by
integration, i.e.,
\begin{equation}
\rho _{\left( \mathrm{sa,}o_{1}o_{2}\right) }(t)=r_{0}\int_{t-\tau }^{t}\rho
_{\left( \mathrm{sa,}o_{1}o_{2}\right) }(t,t^{\prime })dt^{\prime }\text{
with }t-\tau \leqslant t^{\prime }\leqslant t.  \label{B1}
\end{equation}%
Now differentiating both sides of Equation (\ref{B1}), we get $\frac{d\rho
_{\left( \mathrm{sa,}o_{1}o_{2}\right) }(t)}{dt}=r_{0}\frac{d}{dt}%
\int_{t-\tau }^{t}\rho _{\left( \mathrm{sa,}o_{1}o_{2}\right) }(t,t^{\prime
})dt^{\prime }$, which can be expanded by means of the Leibnitz rule as

\begin{equation}
\frac{d\rho _{\left( \mathrm{sa,}o_{1}o_{2}\right) }(t)}{dt}=r_{0}\left[
\rho _{\left( \mathrm{sa,}o_{1}o_{2}\right) }(t,t)-\rho _{\left( \mathrm{sa,}%
o_{1}o_{2}\right) }(t,t-\tau )\right] +r_{0}\int_{t-\tau }^{t}\frac{\partial
\rho _{\left( \mathrm{sa,}o_{1}o_{2}\right) }(t,t^{\prime })}{\partial t}%
dt^{\prime }.  \label{B2}
\end{equation}%
In the Markov approximation \cite{gardiner} where the atomic and laser modes
states are assumed to be uncorrelated at the instant $t$ in which the atoms
are injected in the cavity, we have $\rho _{\left( \mathrm{sa,}%
o_{1}o_{2}\right) }(t,t)=\rho _{o_{1}o_{2}}(t)\rho _{\mathrm{sa}}(0)$.
Besides, we suppose that all the three-level atoms to be initially injected into
the cavity in the lower energy level $|\ell _{3}\rangle $. Then, the
corresponding initial density operator reads $\rho _{\mathrm{sa}%
}(0)=|\ell _{3}\rangle \langle \ell _{3}|$. Furthermore, assuming that the
atomic and laser modes states are uncorrelated just after the atoms left the
cavity, which allows us to write $\rho _{\left( \mathrm{sa,}%
o_{1}o_{2}\right) }(t,t-\tau )=\rho _{o_{1}o_{2}}(t)\rho _{\mathrm{sa}%
}(t-\tau )$, therefore Equation (\ref{B2}) yields
\begin{equation}
\frac{d\rho _{\left( \mathrm{sa,}o_{1}o_{2}\right) }(t)}{dt}=r_{0}\left[
\rho _{\mathrm{sa}}(0)-\rho _{\mathrm{sa}}(t-\tau )\right] \rho
_{o_{1}o_{2}}(t)+r_{0}\int_{t-\tau }^{t}\frac{\partial \rho _{\left( \mathrm{%
sa,}o_{1}o_{2}\right) }(t,t^{\prime })}{\partial t}dt^{\prime },  \label{B3}
\end{equation}%
where $\rho _{\mathrm{sa}}(t-\tau )$ is the density operator for a single atom
pumped at $t-\tau $. Henceforth, we use $\rho _{o_{1}o_{2}}(t)\equiv \rho $
for simplicity of notation.

Obviously, we have $\frac{\partial }{\partial t}\rho _{\left( \mathrm{sa,}%
o_{1}o_{2}\right) }(t)=\frac{-\mathrm{i}}{\hbar }[\mathcal{H}_{af}^{int}+%
\mathcal{H}_{ac}^{int},\rho _{\left( \mathrm{sa,}o_{1}o_{2}\right) }(t)]$
which together with Equations (\ref{B1}) and (\ref{B3}) leads to
\begin{equation}
\frac{d\rho _{\left( \mathrm{sa,}o_{1}o_{2}\right) }(t)}{dt}=r_{0}\left[
\rho _{\mathrm{sa}}(0)-\rho _{\mathrm{sa}}(t-\tau )\right] \rho -\frac{%
\mathrm{i}}{\hbar }[\mathcal{H}_{af}^{int}+\mathcal{H}_{ac}^{int},\rho
_{\left( \mathrm{sa,}o_{1}o_{2}\right) }(t)].  \label{B4}
\end{equation}

Next, taking into account the damping of the two-mode laser $o_{1}$ and $%
o_{2}$ by the vacuum reservoir, and tracing both sides of Equation (\ref{B4}%
) over the atomic variables, we get
\begin{equation}
\frac{d\rho }{dt}=\frac{-\mathrm{i}}{\hbar }\mathrm{Tr}_{\mathrm{sa}}[%
\mathcal{H}_{af}^{int}+\mathcal{H}_{ac}^{int},\rho _{\left( \mathrm{sa,}%
o_{1}o_{2}\right) }(t)]+\sum\limits_{j=1,2}\kappa _{j}\mathcal{L}\left[ c_{j}%
\right] \rho ,  \label{B5}
\end{equation}%
where we used $\mathrm{Tr}_{\mathrm{sa}}\rho _{\mathrm{sa}}(0)=\mathrm{Tr}_{%
\mathrm{sa}}\rho _{\mathrm{sa}}(t-\tau )=1$ and $\mathrm{Tr}_{\mathrm{sa}%
}\rho _{(\mathrm{sa},o_{1}o_{2})}(t)=\rho $, with $\mathcal{L}\left[ c_{j}%
\right] \rho =2c_{j}\rho c_{j}^{\dag }-\{c_{j}^{\dag }c_{j},\rho \}_{+}$
being the Lindblad damping term \cite{milburn}. Replacing $\mathcal{H}%
_{af}^{int}$ and $\mathcal{H}_{ac}^{int}$ by their expressions in Equation (%
\ref{B5}) and tracing out the atomic variables, we find
\begin{eqnarray}
\frac{d\rho }{dt} &=&\varsigma (\rho _{12}c_{1}^{\dag }-c_{1}^{\dag }\rho
_{12}+\rho _{23}c_{2}^{\dag }-c_{2}^{\dag }\rho _{23}+c_{1}\rho _{21}-\rho
_{21}c_{1}+c_{2}\rho _{32}-\rho _{32}c_{2})+  \notag \\
&&\sum\limits_{j=1,2}\kappa _{j}\mathcal{L}\left[ c_{j}\right] \rho ,
\label{B6}
\end{eqnarray}%
where we used the notation $\rho _{mn}=\langle \ell _{m}|\rho _{\left(
\mathrm{sa,}o_{1}o_{2}\right) }|\ell _{n}\rangle $ for $m,n=1,2,3$.

Now, we determine the density operators $\rho _{12}$ and $\rho _{23}$
involved in Equation (\ref{B6}). For this, we multiply Equation (\ref{B4})
on the left by $\langle \ell _{m}|$, and on the right by $|\ell _{n}\rangle $%
, furthermore we assume that the atoms decay to energy levels other than $%
|\ell _{1}\rangle $, $|\ell _{2}\rangle $ or $|\ell _{3}\rangle $ when they
leave the cavity, i.e., $\langle \ell _{m}\left\vert \rho _{\mathrm{sa}%
}(t-\tau )\right\vert \ell _{n}\rangle =0$. Then, we get
\begin{equation}
\frac{d\rho _{mn}}{dt}=r_{0}\langle \ell _{m}\left\vert \rho _{\mathrm{sa}%
}(0)\right\vert \ell _{n}\rangle \rho -\frac{\mathrm{i}}{\hbar }\langle \ell
_{m}|[\mathcal{H}_{af}^{int}+\mathcal{H}_{ac}^{int},\rho _{(\mathrm{sa}%
,o_{1}o_{2})}]|\ell _{n}\rangle -\gamma _{mn}\rho _{mn},  \label{B7}
\end{equation}%
where the coefficients $\gamma _{mn}$ are added to account for the
spontaneous emission and dephasing processes \cite{sz}.

Hence, using the fact that $\rho _{\mathrm{sa}}(0)=|\ell _{3}\rangle \langle
\ell _{3}|$ and the expressions of $\mathcal{H}_{af}^{int}$ and $\mathcal{H}%
_{ac}^{int}$ together with Equation (\ref{B7}), we obtain
\begin{eqnarray}
\frac{d\rho _{11}}{dt} &=&\varsigma (c_{1}\rho _{21}+\rho _{12}c_{1}^{\dag
})-\frac{\Omega }{2}\left( \rho _{13}+\rho _{31}\right) -\gamma _{1}\rho
_{11},  \label{B8} \\
\frac{d\rho _{22}}{dt} &=&-\varsigma (c_{1}^{\dag }\rho _{12}+\rho
_{21}c_{1}-c_{2}\rho _{32}-\rho _{23}c_{2}^{\dag })-\gamma _{2}\rho _{22},
\label{B9} \\
\frac{d\rho _{33}}{dt} &=&r_{0}\rho -\varsigma (c_{2}^{\dag }\rho _{23}+\rho
_{32}c_{2})+\frac{\Omega }{2}\left( \rho _{13}+\rho _{31}\right) -\gamma
_{3}\rho _{33},  \label{B10} \\
\frac{d\rho _{13}}{dt} &=&\varsigma (c_{1}\rho _{23}-\rho _{12}c_{2})-\frac{%
\Omega }{2}\left( \rho _{33}-\rho _{11}\right) -\gamma _{13}\rho _{13},
\label{B11} \\
\frac{d\rho _{12}}{dt} &=&\varsigma (c_{1}\rho _{22}-\rho _{11}c_{1}+\rho
_{13}c_{2}^{\dag })-\frac{\Omega }{2}\rho _{32}-\gamma _{12}\rho _{12},
\label{B12} \\
\frac{d\rho _{32}}{dt} &=&-\varsigma (\rho _{31}c_{1}-\rho _{33}c_{2}^{\dag
}+c_{2}^{\dag }\rho _{22})+\frac{\Omega }{2}\rho _{12}-\gamma _{23}\rho
_{32},  \label{B13}
\end{eqnarray}%
where $\gamma _{j}$ ($j=1,2,3$) refers to the $j$\textrm{th} atomic-level
spontaneous emission decay rate, $\gamma _{13}$ represents the two-photon
dephasing rate, whereas $\gamma _{12}$ and $\gamma _{23}$ denote the
single-photon dephasing rates \cite{Louissel}. From now on, we take for
simplicity $\gamma _{1}=\gamma _{2}=\gamma _{3}=\gamma _{12}=\gamma
_{13}=\gamma _{23}\equiv \gamma $.

Subsequently, in the good cavity limit where $\kappa _{j}\ll \gamma _{j}$,
the atomic states reach the stationary regime much faster than the laser
modes \cite{sz}. Then we can adiabatically eliminate the dynamics of the
atoms by setting the time derivatives in Equations [(\ref{B8})-(\ref{B11})]
to zero. On the other hand, applying the linear approximation \cite%
{sz,Louissel}, one can eliminate the terms proportional to $\varsigma $ in
the same equations. Thusly
\begin{eqnarray}
-\frac{\Omega }{2}\left( \rho _{13}+\rho _{31}\right) -\gamma \rho _{11} &=&0,
\label{B14} \\
\rho _{22} &=&0,  \label{B15} \\
\ r_{0}\rho +\frac{\Omega }{2}\left( \rho _{13}+\rho _{31}\right) -\gamma
\rho _{33} &=&0,  \label{B16} \\
\ -\frac{\Omega }{2}\left( \rho _{33}-\rho _{11}\right) -\gamma \rho _{13}
&=&0.  \label{B17}
\end{eqnarray}

From Equation (\ref{B17}), we could simply verify that $\rho _{13}=\rho _{31}
$, and in view of Equations (\ref{B14}), (\ref{B16}), and (\ref{B17}), we
obtain

\begin{eqnarray}
\rho _{11} &=&\frac{r_{0}\rho \Omega ^{2}}{2\gamma \left( \gamma ^{2}+\Omega
^{2}\right) },  \label{B18} \\
\ \rho _{33} &=&\frac{r_{0}\rho \left( 2\gamma ^{2}+\Omega ^{2}\right) }{%
2\gamma \left( \gamma ^{2}+\Omega ^{2}\right) },  \label{B19} \\
\ \rho _{13} &=&-\frac{r_{0}\rho \Omega }{2\left( \gamma ^{2}+\Omega
^{2}\right) }.  \label{B20}
\end{eqnarray}

Next, injecting Equations (\ref{B15}), (\ref{B18}), (\ref{B19}) and (\ref%
{B20}) in Equations (\ref{B12}) and (\ref{B13}), and performing the
adiabatic elimination once again, we get
\begin{eqnarray}
\rho _{12} &=&\frac{-\varsigma r_{0}\rho }{\left( 4\gamma ^{2}+\Omega
^{2}\right) \left( \gamma ^{2}+\Omega ^{2}\right) }\left( 3\Omega ^{2}c_{1}+%
\frac{\Omega \left( 4\gamma ^{2}+\Omega ^{2}\right) }{\gamma }c_{2}^{\dag
}\right) ,  \label{B21} \\
\text{\ }\rho _{32} &=&\frac{\varsigma r_{0}\rho }{\left( 4\gamma
^{2}+\Omega ^{2}\right) \left( \gamma ^{2}+\Omega ^{2}\right) }\left( \frac{%
-\Omega \left( -2\gamma ^{2}+\Omega ^{2}\right) }{\gamma }c_{1}+\left(
4\gamma ^{2}+\Omega ^{2}\right) c_{2}^{\dag }\right) .  \label{B22}
\end{eqnarray}

Finally, inserting Equations (\ref{B21}) and (\ref{B22}) into Equation (\ref%
{B6}), we arrive at Equation (\ref{hn}).\newline
\newline

\section*{Appendix C: Derivation of Equations
(\protect\ref{E2}) and (\protect\ref{E3})}

In terms of the master equation (\ref{hn}) and the formula $\frac{d\langle
\mathcal{O}\rangle }{dt}=\mathrm{Tr}\left( \frac{d\rho }{dt}\mathcal{O}%
\right) $, we have

\begin{eqnarray}
\frac{d\langle c_{1}\rangle }{dt} &=&\Xi _{11}\mathrm{Tr}\left( 2c_{1}^{\dag
}\rho c_{1}c_{1}-c_{1}c_{1}^{\dag }\rho c_{1}-\rho c_{1}c_{1}^{\dag
}c_{1}\right) +\Xi _{22}\mathrm{Tr}\left( 2c_{2}\rho c_{2}^{\dag
}c_{1}-c_{2}^{\dag }c_{2}\rho c_{1}-\rho c_{2}^{\dag }c_{2}c_{1}\right)
\notag \\
&&-\Xi _{12}\mathrm{Tr}\left( c_{1}c_{2}\rho c_{1}-c_{2}\rho
c_{1}c_{1}\right) -\Xi _{21}\mathrm{Tr}\left( \rho c_{1}c_{2}c_{1}-c_{2}\rho
c_{1}c_{1}\right)  \notag \\
&&-\Xi _{12}\mathrm{Tr}\left( \rho c_{1}^{\dag }c_{2}^{\dag
}c_{1}-c_{1}^{\dag }\rho c_{2}^{\dag }c_{1}\right) -\Xi _{21}\mathrm{Tr}%
\left( c_{1}^{\dag }c_{2}^{\dag }\rho c_{1}-c_{1}^{\dag }\rho c_{2}^{\dag
}c_{1}\right)  \notag \\
&&+\kappa _{1}\mathrm{Tr}\left( 2c_{1}\rho c_{1}^{\dag }c_{1}-c_{1}^{\dag
}c_{1}\rho c_{1}-\rho c_{1}^{\dag }c_{1}c_{1}\right) +\kappa _{2}\mathrm{Tr}%
\left( 2c_{2}\rho c_{2}^{\dag }c_{1}-c_{2}^{\dag }c_{2}\rho c_{1}-\rho
c_{2}^{\dag }c_{2}c_{1}\right) ,  \label{d1}
\end{eqnarray}%
where we employed the linearity property of the trace operator.

Now, we calculate the eight traces, that appear in Equation (\ref{d1}), using the cyclic property of the
trace operator, and the facts that $\left[
c_{k},c_{k^{\prime }}\right] =\left[ c_{k}^{\dag },c_{k^{\prime }}^{\dag }%
\right] =0$ and $\left[ c_{k},c_{k^{\prime }}^{\dag }\right] =\delta
_{kk^{\prime }}$ for $k$, $k^{\prime }=1,2$, i.e.,

\begin{eqnarray}
\mathcal{T}_{1} &=&\Xi _{11}\mathrm{Tr}\left( 2c_{1}^{\dag }\rho
c_{1}c_{1}-c_{1}c_{1}^{\dag }\rho c_{1}-\rho c_{1}c_{1}^{\dag }c_{1}\right) ,
\notag \\
&=&\Xi _{11}\mathrm{Tr}\left( 2\rho c_{1}c_{1}c_{1}^{\dag }-\rho
c_{1}c_{1}c_{1}^{\dag }-\rho c_{1}c_{1}^{\dag }c_{1}\right) ,  \notag \\
&=&\Xi _{11}\mathrm{Tr}\left( \rho c_{1}c_{1}c_{1}^{\dag }-\rho
c_{1}c_{1}^{\dag }c_{1}\right) =\Xi _{11}\mathrm{Tr}\left( \rho c_{1}\left[
c_{1}c_{1}^{\dag }-c_{1}^{\dag }c_{1}\right] \right) ,  \notag \\
&=&\Xi _{11}\mathrm{Tr}\left( \rho c_{1}\right) =\Xi _{11}\langle
c_{1}\rangle .  \label{d2}
\end{eqnarray}

\begin{eqnarray}
\mathcal{T}_{2} &=&\Xi _{22}\mathrm{Tr}\left( 2c_{2}\rho c_{2}^{\dag
}c_{1}-c_{2}^{\dag }c_{2}\rho c_{1}-\rho c_{2}^{\dag }c_{2}c_{1}\right) ,
\notag \\
&=&\Xi _{22}\mathrm{Tr}\left( 2\rho c_{2}^{\dag }c_{1}c_{2}-\rho
c_{1}c_{2}^{\dag }c_{2}-\rho c_{2}^{\dag }c_{2}c_{1}\right) ,  \notag \\
&=&\Xi _{22}\mathrm{Tr}\left( 2\rho c_{2}^{\dag }c_{2}c_{1}-\rho c_{2}^{\dag
}c_{2}c_{1}-\rho c_{2}^{\dag }c_{2}c_{1}\right) =0.  \label{d3}
\end{eqnarray}

\begin{eqnarray}
\mathcal{T}_{3} &=&-\Xi _{12}\mathrm{Tr}\left( c_{1}c_{2}\rho
c_{1}-c_{2}\rho c_{1}c_{1}\right) ,  \notag \\
&=&-\Xi _{12}\mathrm{Tr}\left( \rho c_{1}c_{1}c_{2}-\rho
c_{1}c_{1}c_{2}\right) =0.  \label{d4}
\end{eqnarray}

\begin{eqnarray}
\mathcal{T}_{4} &=&-\Xi _{21}\mathrm{Tr}\left( \rho
c_{1}c_{2}c_{1}-c_{2}\rho c_{1}c_{1}\right) ,  \notag \\
&=&-\Xi _{21}\mathrm{Tr}\left( \rho c_{1}c_{2}c_{1}-\rho
c_{1}c_{1}c_{2}\right) ,  \notag \\
&=&-\Xi _{21}\mathrm{Tr}\left( \rho c_{1}c_{1}c_{2}-\rho
c_{1}c_{1}c_{2}\right) =0.  \label{d5}
\end{eqnarray}

\begin{eqnarray}
\mathcal{T}_{5} &=&-\Xi _{12}\mathrm{Tr}\left( \rho c_{1}^{\dag }c_{2}^{\dag
}c_{1}-c_{1}^{\dag }\rho c_{2}^{\dag }c_{1}\right) ,  \notag \\
&=&-\Xi _{12}\mathrm{Tr}\left( \rho c_{1}^{\dag }c_{2}^{\dag }c_{1}-\rho
c_{2}^{\dag }c_{1}c_{1}^{\dag }\right) ,  \notag \\
&=&-\Xi _{12}\mathrm{Tr}\left( \rho c_{2}^{\dag }\left[ c_{1}^{\dag
}c_{1}-c_{1}c_{1}^{\dag }\right] \right) =\Xi _{12}\mathrm{Tr}\left( \rho
c_{2}^{\dag }\right) ,  \notag \\
&=&\Xi _{12}\langle c_{2}^{\dag }\rangle .  \label{d6}
\end{eqnarray}

\begin{eqnarray}
\mathcal{T}_{6} &=&-\Xi _{21}\mathrm{Tr}\left( c_{1}^{\dag }c_{2}^{\dag
}\rho c_{1}-c_{1}^{\dag }\rho c_{2}^{\dag }c_{1}\right) ,  \notag \\
&=&-\Xi _{21}\mathrm{Tr}\left( \rho c_{1}c_{1}^{\dag }c_{2}^{\dag }-\rho
c_{2}^{\dag }c_{1}c_{1}^{\dag }\right) =0.  \label{d7}
\end{eqnarray}

\begin{eqnarray}
\mathcal{T}_{7} &=&\kappa _{1}\mathrm{Tr}\left( 2c_{1}\rho c_{1}^{\dag
}c_{1}-c_{1}^{\dag }c_{1}\rho c_{1}-\rho c_{1}^{\dag }c_{1}c_{1}\right) ,
\notag \\
&=&\kappa _{1}\mathrm{Tr}\left( 2\rho c_{1}^{\dag }c_{1}c_{1}-\rho
c_{1}c_{1}^{\dag }c_{1}-\rho c_{1}^{\dag }c_{1}c_{1}\right) ,  \notag \\
&=&\kappa _{1}\mathrm{Tr}\left( \rho c_{1}^{\dag }c_{1}c_{1}-\rho
c_{1}c_{1}^{\dag }c_{1}\right) =\kappa _{1}\mathrm{Tr}\left( \rho \left[
c_{1}^{\dag }c_{1}-c_{1}c_{1}^{\dag }\right] c_{1}\right) ,  \notag \\
&=&-\kappa _{1}\mathrm{Tr}\left( \rho c_{1}\right) =-\kappa _{1}\langle
c_{1}\rangle .  \label{d8}
\end{eqnarray}

\begin{eqnarray}
\mathcal{T}_{8} &=&\kappa _{2}\mathrm{Tr}\left( 2c_{2}\rho c_{2}^{\dag
}c_{1}-c_{2}^{\dag }c_{2}\rho c_{1}-\rho c_{2}^{\dag }c_{2}c_{1}\right) ,
\notag \\
&=&\kappa _{2}\mathrm{Tr}\left( 2\rho c_{2}^{\dag }c_{1}c_{2}-\rho
c_{1}c_{2}^{\dag }c_{2}-\rho c_{2}^{\dag }c_{2}c_{1}\right) ,  \notag \\
&=&\kappa _{2}\mathrm{Tr}\left( 2\rho c_{2}^{\dag }c_{2}c_{1}-\rho
c_{2}^{\dag }c_{2}c_{1}-\rho c_{2}^{\dag }c_{2}c_{1}\right) =0.  \label{d9}
\end{eqnarray}

On the basis of Equations [(\ref{d2})-(\ref{d9})], we get

\begin{eqnarray}
\frac{d\langle c_{1}\rangle }{dt} &=&\mathcal{T}_{1}+\mathcal{T}_{2}+%
\mathcal{T}_{3}+\mathcal{T}_{4}+\mathcal{T}_{5}+\mathcal{T}_{6}+\mathcal{T}%
_{7}+\mathcal{T}_{8},  \notag \\
&=&-\kappa _{1}\langle c_{1}\rangle +\Xi _{11}\langle c_{1}\rangle +\Xi
_{12}\langle c_{2}^{\dag }\rangle .  \label{d10}
\end{eqnarray}

Therefore, Equation (\ref{E2}) can be now obtained by removing the bracket from
Equation (\ref{d10}), and adding the noise operator $\digamma _{1}$ with
zero mean value, i.e., $\langle \digamma _{1}\rangle $. Finally, following the same
approach presented in this Appendix, Equation (\ref{E3}) can be established.\\
\\

\section*{Acknowledgements}

The authors gratefully acknowledge computational resources from the 2MS team
faculty of sciences Meknes. The authors acknowledge the PPR2 project:
(MESRSI-CNRST).

\section*{Conflict of Interest}

The authors declare no conflict of interest.

\section*{Data Availability Statement}

Research data are not shared.

\end{document}